\documentclass[11pt,a4paper]{article}
\pdfoutput=1
\usepackage[latin1]{inputenc}
\usepackage{jheppub}
\newcommand{\mrm}[1]{\mathrm{#1}}

\renewcommand{\b}{{\mathrm b}}

\renewcommand{\d}{{\mathrm d}}
\newcommand{\e}{{\mathrm e}}
\newcommand{\f}{{\mathrm f}}
\newcommand{\g}{{\mathrm g}}

\newcommand{\p}{{\mathrm p}}
\newcommand{\q}{{\mathrm q}}
\newcommand{\s}{{\mathrm s}}

\renewcommand{\u}{{\mathrm u}}

\newcommand{\W}{{\mathrm W}}
\newcommand{\Z}{{\mathrm Z}}

\newcommand{\dbar}{\overline{\mathrm d}}
\newcommand{\fbar}{\overline{\mathrm f}}

\newcommand{\qbar}{\overline{\mathrm q}}

\newcommand{\ubar}{\overline{\mathrm u}}

\newcommand{\WZ}{{\W/\Z}}

\newcommand{\as}{\alpha_{\mathrm{s}}}
\newcommand{\aem}{\alpha_{\mathrm{em}}}
\newcommand{\aw}{\alpha_{\mathrm{w}}}
\newcommand{\aeff}{\alpha_{\mathrm{eff}}}
\newcommand{\stw}{\sin^2 \! \theta_W}
\newcommand{\ctw}{\cos^2 \! \theta_W}

\newcommand{\ECM}{E_{\mathrm{cm}}}
\newcommand{\shat}{\hat{s}}
\newcommand{\that}{\hat{t}}
\newcommand{\uhat}{\hat{u}}

\newcommand{\sigmahat}{\hat{\sigma}}

\newcommand{\kT}{k_{\perp}}

\newcommand{\PT}[1]{\p_{\perp\mathrm{#1}}}

\newcommand{\pT}{p_{\perp}}
\newcommand{\pTs}{p^2_{\perp}}
\newcommand{\pTe}{\p_{\perp\mrm{evol}}}
\newcommand{\pTse}{\p^2_{\perp\mrm{evol}}}

\newcommand{\pTsmin}{p^2_{\perp\mathrm{min}}}

\newcommand{\pTsmax}{p^2_{\perp\mathrm{max}}}

{\end{list}}
\newcounter{enumct}

\newlength{\abstwidth}
\begin{document}
%set sloppy attitude to line breaks
\sloppy
 
\pagestyle{empty}
 
\begin{flushright}
LU TP 14-02\\
MCnet-14-01\\
January 2014
\end{flushright}

\vspace{\fill}

\begin{center}
{\LARGE\bf Weak Gauge Boson Radiation\\[2mm] 
in Parton Showers}\\[10mm]
{\Large Jesper R. Christiansen and Torbj\"orn Sj\"ostrand} \\[3mm]
{\it Theoretical High Energy Physics,}\\[1mm]
{\it Department of Astronomy and Theoretical Physics,}\\[1mm]
{\it Lund University,}\\[1mm]
{\it S\"olvegatan 14A,}\\[1mm]
{\it SE-223 62 Lund, Sweden}
\end{center}

\vspace{\fill}

\begin{center}
\begin{minipage}{\abstwidth}
{\bf Abstract}\\[2mm]
The emission of $\W$ and $\Z$ gauge boson is included in a traditional 
QCD + QED shower. The unitarity of the shower algorithm links the
real radiation of the weak gauge bosons to the negative weak virtual
corrections. The shower evolution process leads to a competition
between QCD, QED and weak radiation, and allows for $\W$ and $\Z$ boson
production inside jets. Various effects on LHC physics are studied,
both at low and high transverse momenta, and effects at higher-energy 
hadron colliders are outlined.   

\end{minipage}
\end{center}

\vspace{\fill}

\phantom{dummy}

\clearpage

\pagestyle{plain}
\setcounter{page}{1}

\section{Introduction}

The appearance of high-quality LHC data has been matched by 
high-quality theoretical cross section calculations. One example
is $\WZ + n$ jets, where NLO cross sections are available for 
up to $\W + 5$ jets \cite{Bern:2013gka}. In most of these studies 
the emphasis is on QCD issues, specifically all real and virtual
corrections to the Born-level $\WZ$ production graph are of a QCD
nature. Separately there has been a range of studies concentrating
on weak corrections to processes at lepton and hadron colliders, see
\cite{Kuroda:1990wn, Degrassi:1992ue, Beccaria:1998qe,%
Ciafaloni:2000df,Ciafaloni:2000rp,Denner:2001mn,Melles:2001ye,%
Moretti:2005ut,Moretti:2006ea,Ciafaloni:2006qu,Baur:2006sn,%
Banfi:2007gu,Kuhn:2009nf,Bell:2010gi,Dittmaier:2012kx,%
Stirling:2012ak,Campbell:2013qaa}
for a representative but not exhaustive selection. 
In this article we will concentrate on jet production at the LHC
and other future hadron colliders from this latter weak point of view, 
i.e.\ study weak real and virtual corrections to QCD processes, 
as a complement to the QCD path. Such weak corrections grows like 
$\aw \ln^2(E^2/m^2_{\WZ})$, where $E$ is the energy scale of 
the hard process, and thus become non-negligible at high energies. 

The possibility of large weak virtual corrections was highlighted by 
one set of calculations \cite{Moretti:2005ut,Moretti:2006ea}, which 
gave a jet rate reduced by by up to 30\% at around the LHC kinematical 
limit. This study included both $\mathcal{O}(\aw)$ virtual 
corrections to $\mathcal{O}(\as^2)$ processes and $\mathcal{O}(\as)$ 
ones to $\mathcal{O}(\aw\as)$ ones, however. Here we are only    
interested in the former, which appears to be significantly less
\cite{Baur:2006sn}, but still not negligible.

Cancellation between real and virtual corrections is familiar 
from QCD and QED. The appearance of soft and collinear singularities
for the emission of a gluon or photon are compensated by infinitely 
negative virtual corrections, with only finite $\mathcal{O}(\alpha)$
terms remaining after the cancellation of infinities ($\alpha = \as$
or $\aem$, respectively). In some calculations a fictitious photon 
mass is used to regularize these divergences, rather then the more 
familiar dimensional regularization scheme, but such a mass has to be 
sent to zero at the end of the calculation. For weak calculations
the finite physical $\WZ$ mass guarantees finite real and virtual 
corrections throughout. That is, the negative 
$\mathcal{O}(\aw \ln^2(E^2/m^2_{\WZ}))$ corrections to the 
two-jet rate induced by virtual $\WZ$ loops should be compensated 
by the class of two-jet events with an additional real $\WZ$ in the 
final state. A complication, relative to QCD and QED, 
is that the flavour change of $\W^{\pm}$ leads to Bloch-Nordsieck 
violations \cite{Ciafaloni:2000df}, where the real and virtual 
effects do not fully cancel.  

The finite mass also means that classes of events with or without 
a $\WZ$ are completely separated. This is not only an advantage. 
Consider, for instance, how the character of a high-energy quark jet 
is changed by the possibility of $\WZ$ emissions in addition to the 
conventional $\g/\gamma$ ones. Recall that high-$\pT$ jets at the LHC 
easily can acquire masses around or above the $\WZ$ mass already by 
$\g$ radiation. A $\WZ$ produced inside a jet and decaying hadronically 
may then be rather difficult to distinguish from QCD emissions. 
It is therefore natural to consider strong, electromagnetic and 
weak emissions in one common framework when confronting data. 

Traditionally there exists two possible approaches to describe
multiple emissions: matrix elements (ME) and parton showers (PS).
Formally ME is the correct way to go, but that presupposes that
it is possible to calculate both real and virtual corrections to 
high orders. If not, the ME approach breaks down in the divergent 
soft and collinear limits. Here the PS approach is more sensible,
since it includes Sudakov form factors to restore unitarity.
In recent years a main activity has been to combine the ME and PS
approaches to achieve the best overall precision \cite{Buckley:2011ms}. 

Up until now, showers have only included QCD and QED emissions,
and $\WZ$ production has been viewed solely as a task for the ME
part of the overall description. In this article we extend the 
showering machinery to contain also the emission of the $\W$ and $\Z$ 
gauge bosons, on equal footing with QCD and QED emissions. 
This has some advantages for high-$\pT$ jets, precisely where
$\WZ$ decay products may be hidden in the core, among other quarks 
and gluons. The shower formalism directly couples the real 
emissions to the virtual corrections, by Sudakov factors. It thereby
becomes straightforward to study residual non-cancellation of 
real and virtual corrections as a function of jet selection criteria.
Another advantage is 
that multiple emission of $\WZ$ bosons is a natural part of the 
formalism, even if this only becomes important at very high jet 
energies. In the other extreme, the shower mechanism may also 
be relevant for the production of $\WZ$ plus multijets at lower
$\pT$ scales, both as a test of the shower approach as such and 
as a building block for merging/matching approaches.   

The development of a weak shower formalism faces several challenges.
One such is the $\WZ$ masses, that induce both kinematical and
dynamical complications. These will mainly be overcome by matching 
to several relevant ME expressions, thereby guaranteeing improved 
precision relative to a PS-only based description. 

The new showers are implemented as parts of the \textsc{Pythia}~8
event generator \cite{Sjostrand:2007gs,Sjostrand:2006za}. Thereby 
they can be combined with the existing QCD and QED shower 
implementations, and with all other aspects of the complete 
structure of hadron-collider events. This allows us to study  
the consequences at LHC for $\WZ$ production in general, and 
for the structure of high-$\pT$ jets in particular. 

In Section 2 we develop the shower formalism needed for $\W$ and $\Z$
bosons, including several new aspects. This formalism is validated 
in Section 3. In Section 4 it is then applied to study consequences 
for jets and $\WZ$ production at the LHC. A brief outlook towards 
results for even higher-energy colliders is found in Section 5.
Finally Section 6 provides a summary and outlook.

\section{The weak shower}

In this section we describe how the production of $\WZ + n$ jets 
is handled. To be more precise, the bulk of the study will be 
concerned with $n \geq 2$, i.e.\ from where production of $\WZ$ inside 
a jet becomes possible. The $n = 0, 1$ processes do not have a direct
overlap with QCD jets, and an existing shower formalism is appropriate
to handle them, as will be described further below.

In principle, the introduction of $\WZ$ emission in showers would only 
involve the introduction of two new splitting kernels. In practice,
the large $\WZ$ masses lead to large corrections, both in the kinematics
handling and in the splitting behaviour. In order to provide a 
reasonably accurate description, within the limits of the shower 
approach, several matrix elements are used as templates 
to provide a correct dependence on the $\WZ$ masses.   

Also other problems will appear, that are new relative to the already
existing QCD/QED formalism, notably that the weak force has 
spin-dependent couplings and that the emission of a $\W$ boson 
changes the flavour of the radiating quark. Further, a complete 
description would need to include full $\gamma^*/\Z^0$ interference, 
but in the following these interference terms will be neglected.
That is, for low virtualities a pure $\gamma^*$ is assumed, and for 
higher virtualities a pure $\Z^0$. This should be a good first 
approximation, since the bulk of the shower activity should be 
in the two peak regions.  

\subsection{The basic shower formalism}

The starting point for shower evolution is the DGLAP evolution 
equation, which can be written as 
\begin{equation}
\d\mathcal{P}_{a\to bc} = \frac{\alpha}{2\pi} \, \frac{\d Q^2}{Q^2}
\, P_{a\to bc}(z) ~,
\label{DGLAP}
\end{equation}
with $\alpha = \as$ or $\aem$, $Q^2$ some evolution variable 
expressing the hardness of the $a\to bc$ branching and $z$ the 
energy-momentum sharing between the daughters $b$ and $c$.
Some further azimuthal $\varphi$ dependence may or may not be 
included. The branchings can be of the character $\q \to \q\g$, 
$\g \to \g\g$, $\g \to \q\qbar$, $\f \to \f \gamma$ 
($\f = \q$ or $\ell$), and $\gamma \to \f\fbar$. To this list 
we now want to add $\q \to \q'\W$ and $\q\to\q\Z^0$, including
subsequent decays $\W \to \f \fbar'$ and $\Z \to \f\fbar$.
The $\WZ$ production mechanism is directly comparable with   
that of $\g/\gamma$, whereas the decays happen with unit probability
and therefore are slightly separated in character from the 
corresponding $\g \to \q\qbar$ and $\gamma \to \f\fbar$ ones.
The difference obviously is related to the $\WZ$ being massive and 
the $\g/\gamma$ ones massless. 

To set the coupling notation, consider the case that the $\WZ$
masses are set to zero. Then the evolution equations for a quark 
can be written in a common form
\begin{eqnarray} 
\d\mathcal{P}_{\q\to \q X} & = & \frac{\aeff}{2\pi} \, 
\frac{\d Q^2}{Q^2} \, \frac{1 + z^2}{1 - z}  ~, \label{DGLAPqqX} \\
\aeff & = & \as \, \frac{4}{3} ~~\mrm{for}~~ \q\to\q\g ~,\\
& = & \aem \, e_{\q}^2 ~~\mrm{for}~~ \q\to\q\gamma ~,\\
& = & \frac{\aem}{\stw \ctw} \, (T^3_{\q} - e_{\q} \stw)^2 
~~\mrm{for}~~ \q_L \to \q_L \Z ~, \label{aeffZL} \\
& = & \frac{\aem}{\stw \ctw} \, (e_{\q} \stw)^2 
~~\mrm{for}~~ \q_R \to \q_R \Z ~, \label{aeffZR} \\
& = & \frac{\aem}{2 \stw} \, |V^{\mrm{CKM}}_{\q\q'}|^2 \,  
~~\mrm{for}~~ \q_L \to \q'_L\W ~, \label{aeffWL} \\
& = & 0 ~~\mrm{for}~~ \q_R \to \q'_R\W  \label{aeffWR} ~.
\end{eqnarray}
Here $L/R$ denotes left-/right-handed quarks, 
$T^3_{\q} = \pm 1/2$ for up/down-type quarks, 
and $V^{\mrm{CKM}}$ is the CKM quark mixing matrix.

It will be assumed that the incoming beams are unpolarized,
i.e.\ that incoming fermions equally often are left- as righthanded. 
Since QCD interactions are spin-independent, a left- or righthanded 
helicity is picked at random for each separate fermion line at the 
hard interaction. (Usually this association is unique, but in cases 
like $\u\u \to \u\u$ a choice is made that favours a small scattering 
angle, using $1/\that^2$ and $1/\uhat^2$ as relative weights.) 
Since the gauge-boson emissions preserve helicity 
(for massless fermions), the choice at the hard process is propagated
through the full shower evolution. The emission rate for a single $\WZ$
boson is not affected by this helicity conservation, relative to what
spin-averaged splitting kernels would have given, but the rate of
several $\WZ$ bosons is increased. This is most easily realized for the
$\W$ case, where a first emission fixes the fermion line to be lefthanded,
and a second $\W$ therefore can be emitted twice as often as with a 
spin-averaged branching kernel.

The formalism for FSR and ISR, for the case of massless gauge bosons, 
is outlined in \cite{Sjostrand:2004ef}. A brief summary is as follows,
for the case that on-shell masses can be neglected.

For FSR the evolution variable for branchings $a \to b c$ is 
$\pTse = z (1-z) Q^2$ where $Q^2$ is the off-shell (timelike) virtuality 
of parton $a$. The evolution equation becomes
\begin{equation}
\d\mathcal{P}_a = \frac{\alpha(\pTse)}{2\pi} \, 
\frac{\d \pTse}{\pTse} \, P_{a\to bc}(z) \, \Delta_a(\pTsmax, \pTse) ~,
\label{forwardevol}
\end{equation} 
where $\Delta_a$ is the Sudakov form factor, i.e. the no-emission 
probability from the initial maximal scale $\pTsmax$ down to the 
current $\pTse$ one \cite{Buckley:2011ms}. It is obtained from an
exponentiation of the real-emission probability in such a way that
unitarity is restored: the total probability for parton $a$ to 
branch, or to reach a lower cutoff scale $\pTsmin$ without branching,
adds to unity. A dipole shower \cite{Gustafson:1987rq} approach 
is used to set the kinematics of a branching. That is, for a QCD shower, 
colour is traced in the $N_C \to \infty$ limit, and thus the radiating 
parton $a$ can be associated with a ``recoiler'' $r$ that carries 
the opposite colour. A gluon is split into two possible contributions
by its colour and anticolour, both as a radiator and as a recoiler.
The $a + r$ system preserves its four-momentum in a branching and,
if viewed in its rest frame, the $a$ and $r$ three-momenta are scaled 
down without a change in direction to give a the mass $Q$. In this 
frame $z$ ($1-z$) is the fraction of the modified $a$ energy that 
$b$ ($c$) takes.

For ISR it is most convenient to use backwards evolution 
\cite{Sjostrand:1985xi}, i.e.\ to start at the hard interaction and 
then proceed towards earlier branchings at lower evolution scales.  
That is, the $a\to bc$ branching process is now interpreted as 
parton $b$ becoming ``unbranched'' into $a$. Parton $b$ has a spacelike
virtuality with absolute value $Q^2$, and the evolution variable is
$\pTse = (1-z) Q^2$. The evolution equation now depends on PDF ratios
\begin{equation}
\d\mathcal{P}_b = \frac{\alpha(\pTse)}{2\pi} \, 
\frac{\d \pTse}{\pTse} \, \frac{x_a f_a(x_a, \pTse)}{x_b f_b(x_b, \pTse)} 
\, P_{a\to bc}(z) \, \Delta_b(\pTsmax, \pTse; x_b) ~,
\label{backwardevol}
\end{equation} 
where again the Sudakov form factor is obtained by an exponentiation
of the real-emission expression, to preserve unitarity.
The parton coming in from the other side of the event defines a 
recoiler $r$, such that $z = x_b/x_a = (p_b + p_r)^2/(p_a + p_r)^2$.
With $b$ originally moving parallel with the incoming beam particle
with a fraction $x_b$ of the beam momentum, the branching requires
a redefinition of kinematics such that afterwards parton $a$ is parallel
with the beam and carries a fraction $x_a$. Accordingly, all the 
outgoing partons produced by the $b + r$ collision are boosted and 
rotated to a new frame. 

Both ISR and FSR are evolved downwards in $\pTse$, starting from 
a $\pTsmax$ scale typically set by the hard interaction at the core
of the event. A branching at a given scale sets the maximum for the
continued evolution. At each step all the partons that potentially
could branch must be included in the sum of possibilities.
There are always two incoming partons that can radiate, while the 
number of outgoing ones increases as the evolution proceeds. 

A third component of perturbative parton production is multiparton
interactions (MPI). These can also conveniently be arranged in a 
falling $\pT$ sequence, and by unitarity acquires a ``Sudakov'' 
factor in close analogy with that in showers \cite{Sjostrand:1987su}.
Therefore both ISR, FSR and MPI can be combined in one common sequence
of falling $\pT$ scales \cite{Corke:2010yf}:
\begin{eqnarray}
  \frac{\d \mathcal{P}}{\d \pT}&=&
  \left( \frac{\d\mathcal{P}_{\mrm{MPI}}}{\d \pT} +
  \sum   \frac{\d\mathcal{P}_{\mrm{ISR}}}{\d \pT}  +
  \sum   \frac{\d\mathcal{P}_{\mrm{FSR}}}{\d \pT} 
  \right)  \nonumber \\
   & \times & \exp \left( - \int_{\pT}^{p_{\perp\mrm{max}}}
  \left( \frac{\d\mathcal{P}_{\mrm{MPI}}}{\d \pT'}  +
  \sum   \frac{\d\mathcal{P}_{\mrm{ISR}}}{\d \pT'}  +
  \sum   \frac{\d\mathcal{P}_{\mrm{FSR}}}{\d \pT'} 
  \right) \d \pT' \right) ~,
\label{masterevolve}
\end{eqnarray}
with a combined Sudakov factor. Each MPI gives further incoming and
outgoing partons that can radiate, so the ISR and FSR sums now both 
run over an increasing number of potentially radiating partons.
The decreasing $\pT$ scale can be viewed as an evolution towards 
increasing resolving power; given that the event has a particular 
structure when activity above some $\pT$ scale is resolved, how might 
that picture change when the resolution cutoff is reduced by some 
infinitesimal $\d\pT$? That is, let the ``harder'' features of the 
event set the pattern to which ``softer'' features have to adapt. 
Specifically, energy--momentum conservation effects can be handled 
in a reasonably consistent manner, where the hardest steps almost 
follow the standard rules, whereas soft activity is reduced by the 
competition for energy, mainly between ISR and MPI. 

For massless particles only kinematics variables such as $\pT$
can set the scale. For weak showers the $\WZ$ mass introduces an 
alternative scale, and this opens up for ambiguities.
Consider if a combination such as $\pTse + k m_{\WZ}^2$, with
$k$ as a free parameter, is used as ordering variable for $\WZ$
emission (but otherwise not affecting kinematics). Then an increased
$k$ will shift $\WZ$ emissions to occur earlier in the combined 
evolution, which gives them a competitive advantage relative to 
QCD/QED emissions. We will later study the impact of such 
possible choices.

A key feature for the efficient generator implementation is that the
real and virtual corrections exactly balance, i.e.\ that
eq.~(\ref{masterevolve}) contains exactly the same $\d\mathcal{P}$
expressions in the prefactor and in the Sudakov factor. This holds 
for QCD and QED emissions to leading-log accuracy, and also for 
$\Z^0$ ones, but not for $\W^{\pm}$ emissions, due to the above-mentioned 
Bloch-Nordsieck violations. It comes about by a combination of two
facts. Firstly, a real emission of a  $\W^{\pm}$ in the initial state 
changes the flavour of the hard process, while a $\W^{\pm}$ loop does 
not. Secondly, the incoming state is not isospin invariant, i.e.\ 
the proton is not symmetric between $\u$ and $\d$ quarks, nor between 
other isospin doublets. Together this leads to a mismatch between 
real and virtual Sudakov logarithms, that is not reproduced in our 
implementation. In that sense our results on the reduced rate of 
events without a $\W$ emission are not trustworthy. But only the 
$\q\q' \to \q\q'$ processes with both quarks lefthanded are affected, 
not ones where either quark is righthanded, nor $\q\g \to \q\g$ 
processes \cite{Ciafaloni:2000rp}. Also, real and virtual corrections
cancel for final-state emissions, so only initial-state ones 
contribute. The total error we make on this count therefore is small, 
in particular compared with true NLO corrections beyond our accuracy.

\subsection{Merging generics}

One of the key techniques that will be used in the following is 
matrix-element merging \cite{Bengtsson:1986hr,Norrbin:2000uu,Miu:1998ju}. 
It can be viewed as a precursor to PowHeg 
\cite{Nason:2004rx,Frixione:2007vw}.

In a nutshell the philosophy is the following. Assume a Born cross
section $\sigma_{\mrm{B}}$, usually differential in 
a number of kinematical variables that we do not enumerate here.
The real NLO correction to this is $\d\sigma_{\mrm{R}}$, 
differential in three further kinematical variables, with ratio
$\d K_{\mrm{ME}} = \d\sigma_{\mrm{R}} / \sigma_{\mrm{B}}$. The 
parton-shower approximation also starts from $\sigma_{\mrm{B}}$
and multiplies this with $\d K_{\mrm{PS}}$, which represents 
the shower branching kernel, cf.\ eq.~(\ref{DGLAP}), 
summed over all possible shower branchings from the Born state,
differential in $Q^2$, $z$ and $\varphi$. At this stage the
Sudakov form factor has not yet been introduced. Now ensure that  
$\d K_{\mrm{PS}} \geq \d K_{\mrm{ME}}$ over
all of phase space, which may be automatic or require some adjustment,
e.g.\ by a multiplicative factor. Then begin the evolution of the
shower from a starting scale $Q_{\mrm{max}}^2$ downwards, with a first 
(= ``hardest'') branching at a $Q^2$ distributed according to
\begin{equation}
\d K_{\mrm{PS}}(Q^2, z, \varphi) \, \exp\left( - \int_{Q^2}^{Q_{\mrm{max}}^2}
\d Q^2 \int \d z \int \frac{\d \varphi}{2\pi} \, 
\d K_{\mrm{PS}}(Q^2, z, \varphi) \right) ~.
\label{PSrate}
\end{equation}
Since $\d K_{\mrm{PS}}$ is an overestimate, accept a branching with 
a probability $\d K_{\mrm{ME}} / \d K_{\mrm{PS}}$. This replaces the 
$\d K_{\mrm{PS}}$ prefactor in eq.~(\ref{PSrate}) by $\d K_{\mrm{ME}}$, 
but leaves the Sudakov unchanged. Now use the veto algorithm trick: 
when a $Q^2$ scale is not accepted, set $Q_{\mrm{max}}^2 = Q^2$ and
continue the evolution down from this new maximal scale. This gives
a distribution 
\begin{equation}
\d K_{\mrm{ME}}(Q^2, z, \varphi) \, \exp\left( - \int_{Q^2}^{Q_{\mrm{max}}^2}
\d Q^2 \int \d z \int \frac{\d \varphi}{2\pi} \, 
\d K_{\mrm{ME}}(Q^2, z, \varphi) \right)
\label{MErate}
\end{equation}
(for proof see e.g.\ \cite{Sjostrand:2006za}). Here the dependence on 
the original $\d K_{\mrm{PS}}$ is gone, except that the shower $Q^2$ 
definition is used to set the order in which the phase space is sampled. 
The soft and collinear divergences leads to eqs.~(\ref{PSrate}) and
(\ref{MErate}) being normalized exactly to unity; a first emission
is always found. In practice a lower cutoff $Q_{\mrm{min}}^2$ is always
introduced; if the evolution falls below this scale then an event of 
the Born character is assumed. This preserves unitarity. 
 
This completes the description of ME merging. In PowHeg the hardness
scale is not based on any specific shower, but fills a similar function.
More importantly, to achieve full NLO accuracy, PowHeg normalizes the 
differential cross section in eq.~(\ref{MErate}) to 
$\sigma_{\mrm{B}} + \sigma_{\mrm{V}} + \int \d\sigma_{\mrm{R}}$,
where $\sigma_{\mrm{V}}$ are the virtual corrections, including 
PDF counterterms. We will not here aim for a corresponding NLO 
accuracy, but keep open the possibility to multiply by an overall
``$K$ factor'', which catches the bulk of the NLO effects. 

A simple application of ME merging is $\WZ + 1$ jet, starting from 
the Born $\WZ$ production process \cite{Miu:1998ju}. The
$\q\qbar \to \Z\g$ (or $\q\qbar' \to \W\g$) final state can be
reached by two shower emission histories, which match the $t$- and 
$u$-channel Feynman graphs of the matrix elements. It is found that
$1/2 < \d K_{\mrm{ME}} / \d K_{\mrm{PS}} \leq 1$, so that Monte Carlo 
rejection is straightforward. (The original result was found for an 
evolution in virtuality rather than in $\pTs$, but both give the same 
result since $\d Q^2/Q^2 = \d \pTs/\pTs$ and $z$ is the same.)  
The $\q\g \to \Z\q$ (or $\q\g \to \W\q'$)  process has one shower
history, with a $\g \to\q\qbar$ branching, that corresponds to the
$u$-channel Feynman diagram, while the $s$-channel quark exchange 
diagram has no shower correspondence. In this 
case $1 \leq \d K_{\mrm{ME}} / \d K_{\mrm{PS}} \leq 
(\sqrt{5} - 1)/(2(\sqrt{5} -2)) < 3$, which requires the shower
emission rate to be artificially enhanced for Monte Carlo rejection 
to work. For both processes agreement is found in the $\pT \to 0$
limit, as it should, with increasing discrepancies at larger $\pT$,
but still only by a modest factor right up to the kinematical limit.

It is plausible that the (uncorrected) PS underestimate of the 
$\q\g \to \Z\q$ emission rate at least partly is related to it 
missing one Feynman graph. If so, the shower description of 
$\WZ + \geq 2$~partons can be expected to do even worse, since
it misses out on further diagrams. This is the behaviour observed 
in data \cite{Abazov:2006gs,Chatrchyan:2011ne, Aad:2013ysa}.  
By starting up from QCD $2 \to 2$ processes 
as well, but avoiding doublecounting, it is the hope to bring up
this rate.  

Given that the ME merging approach has been used to select the hardest
emission, a normal shower can be attached for subsequent emissions 
from this scale downwards. Normally these emissions would be based 
on the shower algorithm pure and simple. In some cases it may be 
convenient to use the merging approach also for subsequent emissions,
notably for massive particles in the final state, where the suppression
of collinear radiation may not be fully described by the shower 
\cite{Norrbin:2000uu,Carloni:2011kk}. Although the ME is not the 
correct one for consecutive emissions, it still encodes the suppression 
from mass terms to a reasonable approximation, at least as well as
a shower could. The one modification is to apply it to changed
kinematical conditions, e.g.\ to a gradually decreasing dipole mass 
for FSR. We will come back to this point.

\subsection{Pure final-state emissions}

As a starting point for FSR we consider the simplest possible case, when a 
$\Z$ or $\W$ is radiated in the final state of an $s$-channel process such as 
$\q_0 \qbar_0 \to \g^*(0) \to \q\qbar \to \q(1) \, \qbar(2) \, \Z^0(3)$
(or $\q_0 \qbar_0 \to \g^*(0) \to \q(1) \, \qbar'(2) \, \W(3)$). 
Using CalcHEP \cite{Belyaev:2012qa} for these and subsequent ME
calculations, the matrix element can be written as 
\begin{equation}
\frac{1}{\sigma_0} \, \frac{\d\sigma}{\d x_1 \, \d x_2} 
= \frac{\aeff}{2\pi} \, \left( 
\frac{x_1^2 + x_2^2 + 2 r_3 (x_1 + x_2) + 2r_3^2}{(1 - x_1)(1 - x_2)}
- \frac{r_3}{(1 - x_1)^2} - \frac{r_3}{(1 - x_2)^2} \right) ~. 
\label{eq:MEFSR}
\end{equation}
Here $x_i = 2 p_0 p_i / p_0^2 = 2 E_i/\ECM$, with the latter expression 
valid in the rest frame of the process, and $r_i = m_i^2/\ECM^2$,  
here with the quarks assumed massless. In order to arrive at the 
above result, the ME was integrated over three angular variables.
Setting $r_3 = 0$ the kinematics dependence reverts to the familiar one 
for three-jet events in $\e^+ \e^-$ annihilation, as it should. 
The $\aeff$ values are provided in 
eqs.~(\ref{aeffZL})--(\ref{aeffWR}). 

Owing to the $\WZ$ mass, the phase space for a weak emission is considerably
different from that of a QCD one. Notably the soft and collinear 
divergences lie outside the physical region. Within the accessed region 
we would like to use the matrix-element merging approach to achieve
as accurate a description as possible. As a first step, an 
overestimate is obtained by 
\begin{equation}
\frac{1}{\sigma_0} \, \frac{\d\sigma}{\d x_1 \, \d x_2} 
\leq \frac{\aeff}{2\pi} \, 
\frac{N}{(1 - x_1)(1 - x_2)} ~, 
\end{equation}
with $N = 8$. This translates into an overestimate 
\begin{equation} 
\d\mathcal{P}_{\q\to \q \Z}^{\mrm{over}} = \frac{\aeff}{2\pi} \, 
\frac{\d\pTse}{\pTse} \, \frac{N}{1 - z}  ~,
\label{eq:PqqZFSR}
\end{equation}
which later is to be corrected.

The emission of heavy bosons in final state radiation has already 
been considered in the context of massive Hidden-Valley photons
\cite{Carloni:2011kk}, and therefore only a short review is provided 
here. Consider the process $p_0 \to p_{13} + p_2 \to p_1 + p_2 + p_3$, 
where all particles are allowed to be massive. While the matrix 
elements are described by $x_1$ and $x_2$ (after a suitable integration 
over angles), the parton shower is described in terms of 
\begin{equation}
\pTse = z (1-z) (m_{13}^2 - m_1^2)
\end{equation}
and $z$, which in the massless limit equals $x_1/(x_1 + x_3)$.
For a massive case it is convenient to start out from two massless 
four-vectors $p_1^{(0)}, p_3^{(0)}$ and then create the massive ones 
as linear combinations
\begin{eqnarray}
p_1 & = & (1 - k_1) p_1^{(0)} + k_3 p_3^{(0)} ~, \\
p_3 & = & (1 - k_3) p_3^{(0)} + k_1 p_1^{(0)} ~, \\
k_{1,3} & = & \frac{m_{13}^2 - \lambda_{13}
  \pm (m_3^3 - m_1^2)}{2 m_{13}^2} ~,  \\
\lambda_{13} & = & \sqrt{(m_{13}^2 - m_1^2 - m_3^2)^2 - 4 m_1^2 m_3^2} ~.
\end{eqnarray}
This new energy sharing corresponds to a rescaled
\begin{equation}
z = \frac{1}{1 - k_1 - k_3} \left( \frac{x_1}{2 - x_2} - k_3 \right) ~. 
\label{zFSRresc} 
\end{equation}  
The $\pTse$ and $z$ expressions, can be combined to give the Jacobian
\begin{equation}
\frac{\d \pTse}{\pTse} \, \frac{\d z}{1 - z} = 
\frac{\d x_2}{ 1 - x_2 + r_2 - r_1} \, \,
\frac{\d x_1}{x_3 - k_1 (x_1 + x_3)} ~.  
\end{equation}
Note that the shower expressions so far only referred to emissions 
from the $\q(1)$, whereas the matrix elements also include emissions
from the $\qbar(2)$ and interferences. For a ME/PS comparison it is 
therefore necessary either to sum the two PS possibilities or split 
the ME expression. We choose the latter, with a split in proportion
to the propagators, which gives a probability for the $\q(1)$
\begin{equation}
P_1 = \frac{(m_{13}^2 - m_1^2)^{-1}}{(m_{13}^2 - m_1^2)^{-1} +
(m_{23}^2 - m_2^2)^{-1}} = 
\frac{1 - x_1 + r_1 - r_2}{x_3} ~.
\end{equation}
Thus we arrive at the ME/PS correction factor 
\begin{eqnarray} 
W_1 = \frac{W_{ME,1}}{W_{PS,1}} & = & 
\frac{(1 - x_1 + r_1 - r_2)(1 - x_2 + r_2 - r_1)}{N} \, \,
\frac{x_3 - k_1 (x_1 + x_3)}{x_3} \,  \nonumber \\
& \times & \frac{1}{\sigma_0} \, \frac{\d \sigma}{\d x_1 \,\d x_2} ~. 
\label{weightFSR} 
\end{eqnarray}  
All the explicit dependence on $m_3$ is located in $k_1$ in the 
last factor, but obviously implicitly the whole kinematics setup 
is affected by the value of $m_3$.

The emission of $\W$ bosons introduces flavour changes in the shower, 
and thus also the need for implementing the full CKM-matrix in the 
emissions, eq.~(\ref{aeffWL}). The change of flavour to top is
excluded due to the high mass of the top quark, which significantly
reduces $\W$ emission off $\b$ quarks. All quarks are considered massless 
in the ME weights, but proper masses are included in the kinematics
calculations, as demonstrated above. 

The ME merging technique, viewed as a correction to the LO expression,
is properly valid only for the first branching. The arguments for 
including a sensible behaviour in the soft and collinear regions
remain, however. Therefore eq.~(\ref{weightFSR}) can be applied 
at all steps of the shower evolution. That is, starting from an 
original $\q\qbar$ dipole, the downwards evolution in $\pTse$
gradually reduces the dipole mass by the $\g/\gamma/\W/\Z$ emissions.
When a $\WZ$ is emitted, the ME correction is based on the current
dipole mass and the emission kinematics. This is particularly relevant
since it may be quite common with one or a few QCD emissions before
a $\WZ$ is emitted.  

In non-Abelian theories the radiated gauge bosons themselves carry 
charge and can themselves radiate. For QCD emissions this is well
approximated by the dipole picture, where each emission of a further
gluon leads to the creation of a new dipole. Similarly the emission
of a $\WZ$ leads to more weak charges, with the possibility of non-Abelian 
branchings $\W^{\pm} \to \W^{\pm} \Z^0$ and $\Z^0 \to \W^+\W^-$. 
So far we have not included these further branchings, and therefore
preserve the original $\q\qbar$ weak-dipole when a $\WZ$
is emitted. This will imply some underestimation of multiple-$\WZ$ 
production rate.

New $\q\qbar$ pairs can be created within the shower evolution, 
e.g.\ by gluon branchings $\g\to\q\qbar$. These are considered as 
new weak dipoles, and can thus enhance the rate of $\WZ$ emissions.

\subsection{Pure initial-state emissions}

As a starting point for ISR we here instead consider a process such as
$\q(1) \, \qbar(2) \to \Z(3) \, \g^*(4)$ 
(or $\q(1) \, \qbar'(2) \to \W(3) \, \g^*(4)$), where the subsequent 
$\g^* \to \q_0\qbar_0$ (or $\g^* \to \g \g$) decay has been integrated 
out. This matrix element can then be written as 
\begin{equation}
W_{\mrm{ME}} = \frac{\shat}{\sigmahat_0} \frac{\d\sigmahat}{\d\that} = 
\frac{\aeff}{2\pi} \, \left( 
\frac{\that^2 + \uhat^2 + 2\shat(m_3^2 + m_4^2)}{\that\uhat}
- \frac{m_3^2 m_4^2}{\that^2} - \frac{m_3^2 m_4^2}{\uhat^2} \right)  ~.
\label{MEISR}
\end{equation}

The ISR kinematics is already set up to handle the emission of a massive
particle, e.g.\ in $\b \to \g\b$, with a $\b$ quark in the final state.
The ME correction machinery \cite{Miu:1998ju} has only been set up for
the emission of a massless particle, however, so some slight changes
are necessary. For the case that the $\WZ$ is emitted by the incoming
parton 1 the Mandelstam variables become
\begin{eqnarray} 
\shat & = & (p_1 + p_2)^2 = \frac{m_4^2}{z} ~, \\
\that & = & (p_1 - p_3)^2 = - Q^2 = - \frac{\pTse}{1 - z} ~, \\
\uhat & = & m_3^2 + m_4^2 - \shat - \that =  m_3^2 + m_4^2 - \frac{m_4^2}{z}
 + \frac{\pTse}{1 - z} ~.  
\end{eqnarray}  
It turns out that the massless DGLAP-kernel eq.~(\ref{DGLAPqqX}) is not an
overestimate for the ME eq.~(\ref{MEISR}). Instead the following slightly
modified splitting kernel is used
\begin{equation}
  \d\mathcal{P}_{\q\to \q X} = \frac{\alpha_{\mrm{eff}}}{2\pi} \, 
  \frac{\d Q^2}{Q^2} \, \frac{1 + z^2(1+r^2)^2}{1 - z(1+r^2)} \, 
  \label{eq:PSISR}
\end{equation}
where $r = m_3 / m_{\mrm{dipole}} = m_3/m_4$. The standard 
DGLAP kernel is recovered in the massless limit. Using the Jacobian 
$\d\that/\that = \d\pTse/\pTse$, the shower emission rate translates to 
\begin{equation}
W_{\mrm{PS1}} = \frac{\shat}{\sigmahat_0} \frac{\d\sigmahat}{\d\that} 
 = \frac{\aeff}{2\pi} \, 
\frac{\shat^2 + (m_3^2 + m_4^2)^2}{\that(\that + \uhat)} ~.
\end{equation} 
Adding the emission from parton 2, easily obtained by 
$\that \leftrightarrow \uhat$, gives
\begin{equation}
W_{\mrm{PS}} = W_{\mrm{PS1}} + W_{\mrm{PS2}} =
\frac{\aeff}{2\pi} \, \frac{\shat^2 + (m_3^2 + m_4^2)^2}{\that\uhat} ~.
\end{equation}
In this case it is convenient to use $W = W_{\mrm{ME}} / W_{\mrm{PS}}$
as ME correction factor. That is, the full ME is compared with the 
sum of the two PS possibilities, unlike the FSR case, where the ME 
is more easily split into two parts each compared with a single 
shower history.

It can most easily be seen that the modified DGLAP kernel is an 
upper estimate by taking the ratio of the PS weight with the ME one, 
\begin{eqnarray}
  W = \frac{W_{\mrm{ME}}}{W_{\mrm{PS}}} & \leq & 
  \frac{\that^2 + \uhat^2 + 2\shat(m_3^2 + m_4^2)}{\shat^2 + (m_3^2+m_4^2)^2}
  \\
  & = & \frac{\that^2 + \uhat^2 + 2\shat^2 + 2\shat\that +
    2\shat\uhat}{\that^2 + \uhat^2 + 2\shat^2 + 2\shat\that 
    + 2\shat\uhat + 2\that \uhat} \leq 1
\end{eqnarray}

A new upper estimate for the range of allowed $z$ values is needed, 
since the standard one enters unphysical regions of the modified DGLAP 
kernel, turning the PS weight negative. This is not a surprise, since 
the standard upper estimate does not include massive emissions. The upper 
estimate chosen is 
\begin{equation}
  z \leq \frac{1}{1+r^2+\frac{\pTse}{m_\mrm{dipole}^2}}
\end{equation}
This limit should ensure that the emitted particle will always have 
enough energy to become massive and have the chosen $\pTse$. It is 
not formally proven to be an upper limit, but works for all studied
cases.

The handling of CKM weights for $\W$ emission becomes slightly more 
complicated in ISR than in FSR, owing to the presence of PDFs in 
the evolution. The PDF ratio in eq.~(\ref{backwardevol}) is 
generalized to an upper estimate 
\begin{equation}
R_{\mrm{max}}^{\mrm{PDF}} = \frac{\sum_a |V_{ab}^{\mrm{CKM}}|^2 \, 
x_b f_a(x_b, \pTsmax)}{x_b f_b(x_b, \pTsmax)} 
\end{equation}
used in the downwards evolution with the veto algorithm. For a trial 
emission the relevant part of the acceptance weight then becomes 
\begin{equation}
\frac{1}{R_{\mrm{max}}^{\mrm{PDF}}} \, \frac{\sum_a |V_{ab}^{\mrm{CKM}}|^2
\, x_a f_a(x_a, \pTse)}{x_b f_b(x_b, \pTse)} ~. 
\end{equation}
Once a branching has been accepted, the new mother flavour $a$ is 
selected in proportion to the terms in the numerator sum. 

Like for final-state radiation, the ME merging weight will be used
not only for a $\WZ$ emission in direct association with the hard
process, but for all branchings in the backwards evolution.
All final-state particles are then lumped into one single effective
particle, like the $\g^*$ above.

\subsection{Mixed initial--final-state emissions}

In addition to the pure-final or pure-initial topologies, the two other
relevant possibilities are with one or two quark lines flowing through
the hard $2 \to 2$ topologies, i.e.\ $\q\g \to \q\g$ and 
$\q\q' \to \q\q'$.

It would have been tempting to use the ME correction factors as above 
for FSR and ISR. Unfortunately this does not give a particularly good
agreement with the $\q\g \to \q\g\Z^0$ matrix element. Specifically,
whereas $s$-channel processes tend to populate the available phase space
with only a $\d\pTs/\pTs$ fall-off, the coherence between ISR and FSR 
in $t$-channel processes leads to a destructive interference that 
reduces emissions at large angles \cite{Ellis:1986bv}. Thus emission
rates depend on the $\that$  of the core $2 \to 2$ process, 
not only on its $\shat$. Therefore we have chosen to base the ME 
corrections on the full $2 \to 3$ kinematics.

The general strategy will be to use that the three-body phase space 
can be split into two two-body ones, with an intermediate state $i$,
e.g.
\begin{equation}
\d\Phi_3(1 + 2 + 3) = \d\Phi_2(1 + i) \, \frac{\d m_i^2}{2\pi} \, 
\d\Phi_2(i \to 2 + 3) ~.
\end{equation}
One of the $\d\Phi_2$ factors will be associated with the QCD hard
$2 \to 2$ process, whereas the rest comes from the shower branching.
This way it is possible to compare the $2 \to 3$ ME with the 
$2 \to 2$ ME + shower in the same phase space point, with proper
Jacobians. 

To begin with, consider the simpler first process, $\q\g \to \q\g$,
with an additional $\Z^0$ emission, labeled as 
$\q(a) \, \g(b) \to \q(1) \, \g(2) \, \Z^0(3)$. We will first outline
the procedures for FSR and ISR separately, and then explain how to
combine the two, and how to modify for $\W^{\pm}$ emission.

For FSR the intermediate state is the virtual quark that emits the $\Z^0$,
$\q(a) \, \g(b) \to \q^*(i) \, \g(2) \to \q(1) \, \g(2) \, \Z^0(3)$, 
which gives the phase space separation
\begin{equation}
\d\Phi_3(a + b \to 1 + 2 + 3) = \d\Phi_2(a + b \to i + 2) \,
\frac{\d m_i^2}{2\pi} \, \d\Phi_2(i \to 1 + 3) ~.
\end{equation}
Rewriting the second $\d\Phi_2$ in terms of angles in the $i$ rest frame, 
the $2 \to 3$ ME can be expressed as
\begin{equation}
\d\sigma_{\mrm{ME}} = \frac{|M_{2 \to 3}|^2}{2\shat} \, \d\Phi_3
= \frac{|M_{2 \to 3}|^2}{2\shat} \, \d\Phi_2(i + 2) \, \frac{\d m_i^2}{2\pi} 
\, \frac{\beta_{13}}{4} \, \d(\cos\theta^*) \,
\frac{\d\varphi^*}{2\pi} ~,
\end{equation}
with
\begin{equation}
\beta_{jk} = \sqrt{ \left( 1 - \frac{m_j^2}{m_{jk}^2} - \frac{m_k^2}{m_{jk}^2} 
\right)^2 - 4  \frac{m_j^2}{m_{jk}^2} \frac{m_k^2}{m_{jk}^2} } ~,~~
m_{jk}^2 = (p_j + p_k)^2 ~,
\end{equation}
which simplifies to $\beta_{13} = 1 - m_3^2/m_i^2$ if $m_1 = 0$.

The $2 \to 2$ ME combined with the shower instead gives an answer
\begin{equation}
\d\sigma_{\mrm{PS}} = \frac{|M_{2 \to 2}|^2}{2\shat} \, \d\Phi'_2(i + 2)
\, \frac{\aeff}{2\pi} \, \frac{\d\pTse}{\pTse} \, 
\frac{N \, \d z}{1 - z} \, \frac{\d\varphi^*}{2\pi} ~. 
\end{equation}
Here $\d\Phi'_2(i + 2)$ represents the outgoing $i$ before it acquires 
a mass by the $\q^* \to \q\Z^0$ branching, as assumed for the initial 
$2 \to 2$ QCD process. The correct phase space, used in the ME expression,
is scaled down by a factor $\beta_{i2} = 1 - m_i^2/\shat$. 
To compare the two rates, it is necessary
to convert between the standard two-body phase-space variables and 
the shower ones. The relationship $\pTse = z(1-z) m_i^2$ gives
$\d\pTse / \pTse = \d m_i^2 / m_i^2$. For $z$ it is convenient to define 
kinematics in the $i$ rest frame, $p_i^* = m_i(1; 0, 0, 0)$, with 2 
along the $-z$ axis. Then, with $m_1 = 0$,
\begin{eqnarray}
p^*_1 & = & \frac{m_i^2 - m_3^2}{2m_i} \,
\left( 1; \sin\theta^*, 0, \cos\theta^* \right) ~, \\
p^*_0 & = & p^*_1 + p^*_2 + p^*_3 = \left( \frac{\shat + m_i^2}{2m_i} ;
0, 0, - \frac{\shat - m_i^2}{2m_i} \right) ~.
\end{eqnarray}
Now insert into eq.~(\ref{zFSRresc}), with $k_1 = m_3^2 / m_i^2$
and $k_3 = 0$,
\begin{equation}
z = \frac{1}{1 - m_3^2/m_i^2} \, \frac{x_1}{x_i}
 = \frac{m_i^2}{m_i^2 - m_3^2} \, \frac{p^*_0 p^*_1}{p^*_0 p^*_i}
 = \frac{1}{2} \left( 1 + \frac{\shat - m_i^2}{\shat + m_i^2} 
   \, \cos\theta^* \right) ~,
\end{equation}
from which $\d(\cos\theta^*) / \d z$ can be read off. This gives a 
ME correction weight to the shower
\begin{eqnarray}
W_{\mrm{FSR}}  & = & \frac{\d\sigma_{\mrm{ME}}}{\d\sigma_{\mrm{PS}}}  
= \frac{|M_{2 \to 3}|^2 \, \d\Phi_2(i + 2)}{|M_{2 \to 2}|^2 \, \d\Phi'_2(i + 2)} 
\, \frac{\beta_{13}}{4 \, \aeff \, N}
\frac{ \pTse \, \d m_i^2}{\d \pTse} \, (1 - z) \, 
\frac{\d(\cos\theta^*)}{\d z} \nonumber \\
 & = & \frac{|M_{2 \to 3}|^2}{|M_{2 \to 2}|^2} \, \beta_{i2} 
\frac{\beta_{13}}{2 \, \aeff \, N} \, m_i^2 \, (1-z) \, 
\frac {\shat + m_i^2}{\shat - m_i^2} \nonumber \\
 & = & \frac{|M_{2 \to 3}|^2}{|M_{2 \to 2}|^2} \, 
\frac{1}{2 \, \aeff \, N} \, \frac{\pTse}{z} \, 
\frac {\shat}{\shat - m_i^2} \, \frac{m_i^2 - m_3^2}{m_i^2} ~.
\end{eqnarray}

For ISR the intermediate state instead is the $2\to 2$ QCD process
$\q(a) \, \g(b) \to (\q^*\g)(i) \, \Z^0(3) \to \q(1) \, \g(2) \, \Z^0(3)$, 
where the $\q^*$ is the spacelike quark after having emitted the $\Z^0$.
Thus the phase space separation here is 
\begin{equation}
\d\Phi_3(a + b \to 1 + 2 + 3) = \d\Phi_2(a + b \to i + 3) \,
\frac{\d m_i^2}{2\pi} \, \d\Phi_2(i \to 1 + 2) ~.
\end{equation}
The first $\d\Phi_2$ is rewritten in terms of angles in the $a + b$ 
rest frame, giving 
\begin{equation}
\d\sigma_{\mrm{ME}} = \frac{|M_{2 \to 3}|^2}{2\shat} \, \d\Phi_3
= \frac{|M_{2 \to 3}|^2}{2\shat} \, \frac{\beta_{i3}}{4} \, \d(\cos\theta) 
\, \frac{\d\varphi}{2\pi} \, \frac{\d m_i^2}{2\pi} \, \d\Phi_2(1 + 2) ~,
\end{equation}
while the shower gives
\begin{equation}
\d\sigma_{\mrm{PS}} =  \frac{\aeff}{2\pi} \, 
\frac{\d\pTse}{\pTse} \, \frac{(1+z^2(1+r^2)^2) \, \d z}{1 - z(1+r^2)} \, 
\frac{\d\varphi}{2\pi} \, \frac{|M_{2 \to 2}|^2}{2 m_i^2} \, 
\d\Phi_2(1 + 2) ~. 
\end{equation}
The relation $m_i^2 = z \shat$ gives $\d m_i^2 / \d  z = \shat$.
To relate $\cos\theta$ and $\pTse$ it is convenient to go via
the spacelike virtuality $Q^2$ of the $\q^*$ propagator, which by 
definition is related as $\pTse = (1-z) Q^2$. In the rest frame,
$p_{a,b} = (\sqrt{\shat}/2) \, (1; 0, 0, \pm 1)$, $p_3$ can be written as
\begin{equation}
p_3 = \frac{\sqrt{\shat}}{2} \, \left( \frac{\shat + m_3^2 - m_i^2}{\shat} ;
- \beta_{i3}\sin\theta, 0, - \beta_{i3}\cos\theta \right) ~,
\end{equation}
and thus 
\begin{equation}
Q^2 = - (p_a - p_3)^2 = \frac{1}{2} \left( \shat - m_3^2 - m_i^2 
+ \shat \beta_{i3}\cos\theta \right) ~,
\end{equation}
i.e.\ $\shat \beta_{3i} \d(\cos\theta) / \d\pTse = 2 / (1 - z)$.
Put together, this gives
\begin{eqnarray}
W_{\mrm{ISR}} & = & \frac{\d\sigma_{\mrm{ME}}}{\d\sigma_{\mrm{PS}}}  
= \frac{\frac{|M_{2 \to 3}|^2}{2\shat} \, \frac{\beta_{i3}}{4} \, 
\d(\cos\theta) \, \frac{\d m_i^2}{2\pi}}{\frac{|M_{2 \to 2}|^2}{2 m_i^2}
\frac{\aeff}{2\pi} \, \frac{\d\pTse}{\pTse} \, 
\frac{(1+z^2(1+r^2)^2) \, \d z}{1 - z(1+r^2)}}  \nonumber \\
 & = & \frac{|M_{2 \to 3}|^2}{|M_{2 \to 2}|^2} \,
\frac{\shat \beta_{i3}\d(\cos\theta)}{\d\pTse} \,
\frac{z \pTse}{4 \aeff } \frac{1 - z(1+r^2)}{1+z^2(1+r^2)^2} \nonumber \\
 & = & \frac{|M_{2 \to 3}|^2}{|M_{2 \to 2}|^2} \, \frac{1}{2\aeff} \,
\,\frac{z \pTse(1 - z(1+r^2))}{(1 - z)(1+z^2(1+r^2)^2)} ~. 
\end{eqnarray}

Two further aspects need to be considered. Firstly, the $2 \to 3$ ME
expression should be compared with the sum of the FSR and ISR 
contributions. This could become tedious, so here a simpler route is
to split the ME into two parts, one that is used for the FSR reweighting, 
and another for the ISR one. The relative fractions are chosen by
the respective propagator, which gives an additional factor 
\begin{equation}
W_{\mrm{split, FSR}} 
= \frac{m^{-2}_{i\mrm{(FSR)}}}{m^{-2}_{i\mrm{(FSR)}} + Q^{-2}_{\mrm{(ISR)}}} 
= \frac{|(p_a - p_3)^2|}{|(p_a - p_3)^2| + (p_1 + p_3)^2}
= 1 - W_{\mrm{split, ISR}} ~. 
\end{equation}

Secondly, there are some differences for $\W$ emission. As in 
the $s$-channel case, 
the ISR has to include CKM-weighted PDFs and choices of incoming 
flavour. The flavours in the hard process are also different for 
ISR and FSR: a process like $\u\g \to \d\g\W^+$ has a QCD subprocess 
$\d\g \to \d\g$ for ISR and $\u\g \to \u\g$ for FSR. Since QCD is 
flavour-blind, and the MEs are for massless quarks, this is only 
a matter of bookkeeping. 

The matrix elements for processes like $\q\q' \to \q\q'\Z^0$ and
$\q\qbar' \to \q\qbar'\Z^0$, $\q' \neq \q$, are pure $t$-channel.
They therefore have a somewhat different structure from the 
$\q\g \to \q\g\Z^0$ ones. The general pattern from four radiating 
partons can be quite complex, so for the purpose of a correction 
to the parton shower we have chosen to neglect the cross-terms 
between emission from the $\q$ and $\q'$ flavour lines. That is, 
the $2 \to 3$ ME used for correcting emissions off the $\q$ flavour 
line is obtained by letting couplings to the $\q'$ line vanish. 
As we will show later on, this is a reasonable approximation. 
From there on, the procedure is as for $\q\g \to \q\g\Z^0$. That is, 
the remaining ME is split into one part associated with FSR and 
another with ISR. For each of them a correction from PS to ME is done 
using either $W_{\mrm{FSR}}$ or $W_{\mrm{ISR}}$.

For $\q\q\to\q\q\Z^0$ it is not possible to separate by couplings. 
Instead the fermion lines are picked probabilistically with equal probability 
for each combination. Thereafter each line is considered as in the $\q\q' \to
\q\q'\Z^0$ case.

Finally, a $\q\qbar \to \q\qbar$ process is handled as pure 
$s$-channel, just like a $\q\qbar \to \q'\qbar'$ process.

The description so far has been formulated in terms of corrections
to a $\WZ$ emission as the first branching attached to a $2 \to 2$ 
QCD process, i.e. what the matrix elements have been calculated for.
But for it to be useful, the corrections must be applicable for 
emissions at any stage of the shower, i.e. following a number of
earlier QCD, QED and weak emissions. To do that, the whole system 
is converted to a pseudo $2 \to 2$ process, for which the ME 
correction procedure can be applied as above. In particular, this 
should guarantee a proper account of $\WZ$ mass effects.

For FSR, a recoiler is always chosen in the final state. For a 
process like $\q\q' \to \q\q'$ the initial $\q'$ flavour is considered 
as recoiler to $\q$, however many branchings have occurred. For 
$\q\g \to \q\g$, in a consecutive branching $\g \to \g_1\g_2$ 
the new recoiler is chosen to be the one of $\g_1$ and $\g_2$ that
forms the largest invariant mass together with $\q$. The kinematics 
of the branching process is first boosted longitudinally to the 
rest frame of the two incoming partons of the original $2 \to 2$
process, and thereafter boosted to the rest frame of the 
radiator + recoiler. The momenta of the two incoming partons,
still along the beam axis, are rescaled (down) to the same invariant
mass as the outgoing state. Thus a consistent $2 \to 2 \to 3$  
kinematics is arrived at, and ME corrections can applied to this 
sequence as before.

For ISR there is always a unique recoiler, given by the opposite
incoming parton. In this case a core $2 \to 2$ process is constructed
in its rest frame, with incoming partons that need not agree with
the original ones, while the original outgoing partons are scaled
(up) to the same invariant mass. Thus the scattering angle is
preserved, in some sense. The relevant $\Z$ emission is then added on 
to this kinematics, and the ME correction weight can be found.   

\subsection{Doublecounting with weak Born processes}

Throughout the description, doublecounting issues have appeared.
The $2 \to 3$ ME has been split into two parts, one used for the 
FSR ME corrections, and the other for the corresponding ISR ones.
Within either of ISR or FSR, the possibility of radiation from 
two incoming or two outgoing partons is also taken into account.
There remains one significant source of doublecounting, however,
namely the production of a $\WZ$ as a Born process, followed by 
further QCD emissions. That is, starting from $\q\qbar \to \Z^0$,
first-order topologies $\q \qbar \to \g \Z^0$ and 
$\q \g \to \q \Z^0$ will be generated, and from those
$\q \qbar \to \g \g \Z^0$, $\q \qbar \to \q' \qbar' \Z^0$,
$\q \g \to \q \g \Z^0$ and $\g \g \to \q \qbar \Z^0$. 
It is therefore possible to arrive at the same set of 
$2 \to 3$ processes either from a weak or a QCD base process, 
which opens up for another type of doublecounting.

The two production paths, here denoted ``weak'' or ``QCD'' by the
base process, are expected preferentially to populate different 
phase space regions. To begin with, consider only ISR emission, 
and recall that branchings are ordered in $\pTe$, which approximately 
translates into ordering in ordinary $\pT$. In the weak path, the 
$\Z^0$ and its recoiling parton therefore are produced at a larger 
$\pT$ scale than the further parton produced by the next PS branching. 
By contrast, in the QCD path the $\Z^0$ will be added at the lower 
$\pT$. Similarly, FSR in the weak path allows one parton to split 
into two preferentially nearby partons, which thereby both tend to 
be opposite to the $\Z^0$, while FSR in the weak path would 
preferentially place the $\Z^0$ close to either outgoing parton. 

What complicates the picture above is the use of ME corrections
for the QCD path, which are there to include $\WZ$ mass effects and 
ISR/FSR interference, but as a consequence also weights up the 
singular regions associated with the weak path. This makes the
doublecounting issue more severe than if either path only had 
non-singular tails stretching into the singular region of the 
other path. As a technical side effect, the Monte Carlo efficiency 
of the QCD path elsewhere can become low, since the upper limit for 
the ME/PS ratio, needed to set the trial sampling rate, becomes larger 
the closer to the ``unexpected'' singularities the QCD path is 
allowed to come. By contrast, the \textsc{Pythia} description of 
$\WZ$ production only performs ME corrections for the first emission, 
as already discussed, so the weak path is not corrected by any 
$2 \to 3$ MEs.

The solution we have adopted to this issue is to separate the full 
$2 \to 3$ phase space into two non-overlapping regions, in the spirit
of the $\kT$ clustering algorithm \cite{Catani:1993hr,Salam:2009jx}.
That is, for a $2 \to 3$ phase-space point define distances 
\begin{eqnarray}
d_{iB} & = & p_{\perp i}^2 ~, \\
d_{ij} & = & \min \left( p_{\perp i}^2, p_{\perp j}^2 \right) \, \frac{1}{R^2}
\, \left( (y_i - y_j)^2 + (\varphi_i - \varphi_j)^2 \right) ~, 
\end{eqnarray}
that represent the relative closeness to the ISR and FSR singularities,
respectively, with $R$ providing the relative normalization of the two.
Then find the smallest of these distances, disregarding $d_{ij}$ 
combinations that are not associated with ME singularities, such as 
$\Z^0\g$ or $\q\q$. Associate the phase-space point with the weak path 
if a parton is closest to the beam or two partons closest to each other, 
and with the QCD path if the $\Z^0$ is closest to the beam or to a quark.

Starting from weak production, this means that a check is made after 
the shower has emitted two partons, and if the phase-space point lies
in the QCD-path region the event is rejected. Events with at most one
branching thus are not affected, and emissions subsequent to the first 
two are not checked any further. Starting from a QCD event, the 
emission of a $Z^0$ is vetoed if it falls in the weak-path region.
Not much should be made of the asymmetric treatment, in one case the
veto of the whole event and in the other only of the branching:
were it not for the ME correction on the QCD path then neither 
path would populate the ``wrong'' region to any appreciable extent.
The weak-path choice is motivated by starting out from a 
$\q\qbar \to \Z^0$ cross section that is inclusive, so that the 
addition of the QCD path should be viewed as swapping in a better
description of a region that already is covered. A corresponding 
argument for the QCD-path evolution is less obvious, and it is 
simpler to operate as if $\Z^0$ emissions into the wrong region
do not form a part of the shower evolution. 

\subsection{Other shower aspects}

In the description so far the choice of $\WZ$ mass has not been mentioned.
The practical implementation is such that a new $\WZ$ mass is chosen each 
time a trial $\WZ$ emission is to be defined, according to a relativistic
Breit-Wigner with running width. This allows the $\WZ$ mass distribution 
to be reweighted by the mass dependence of matrix elements and of 
phase space.

In addition, by default there is a lower cutoff at 10 GeV for the $\WZ$
mass. This is intended to avoid doublecounting between the PS description
of $\gamma^*$ production below this scale and the ME description of 
$\gamma^*/\Z^0$ production above it. For the purposes of this study 
the contribution below 10 GeV is negligible. More relevant is the 
absence of the $\gamma^*$ contribution above 10 GeV, and the 
$\gamma^*/\Z^0$ interference contribution everywhere. This could become 
a further extension some day, but would involve relatively minor 
corrections to the overall picture, which is dominated by the $\WZ$ peak
regions.

The emitted weak bosons are decayed after the evolution of the full 
parton shower, into the possible decay channels according to their
partial widths. In order to achieve a better description of the decay 
angles, a ME correction is applied. For FSR this is corrected to the 
ME of a four-body final state, e.g.\ 
$\g^* \to \u\ubar \to \u\ubar\Z \to \u\ubar\e^+\e^-$. The ME is 
based on the helicity previously chosen for the radiating fermion line. 
Since the weak boson is already produced, all overall factors that 
do not depend on the decay angles are irrelevant, including the $\WZ$
propagators. An upper estimate of the ME expression is obtained by
taking four times the maximum obtained for six different decay directions 
($\pm \hat{x}, \pm \hat{y}, \pm \hat{z}$ in the $\WZ$ rest frame);
empirically this always works. Then standard hit-and-miss Monte Carlo
is used to pick the decay direction. For ISR the same method is applied, the
only difference is the change of ME to the $\u \ubar \to \g^* \Z \to \g^*
\e^+\e^-$. In the case of the mixed-initial-final state, the same two MEs are
applied and the choice between them is made depending on where in the shower
algorithm the emission is produced.

After the decay of the weak boson, a new parton shower is started, 
with the two decay products defining the first dipole.

The implementation of the weak shower only works for $2 \to 2$ or $2 \to 1$
hard processes. The reason behind this is that the mixed initial and final
state ME correction relies on a $2 \to 2$ hard process. And if the
starting point would be a $2 \to 3$ process, it is not always possible to 
identify a unique $2 \to 2$ process. 

\section{Validation}

In this section we collect some checks on the functioning of the 
weak-shower implementation. This provides insight into the 
approximations made and their limitations. Needless to say, 
many further checks have been made.

\subsection{Control that parton showers offer overestimates}

As the implementation relies heavily on correcting the shower behaviour 
by a ME/PS ratio, it is relevant to study the correction procedures.
Specifically, the uncorrected PS should everywhere have a higher 
differential rate than the corresponding ME-corrected one has.

\begin{figure}[tbp]
  \centering
  \includegraphics[width=0.49\textwidth]{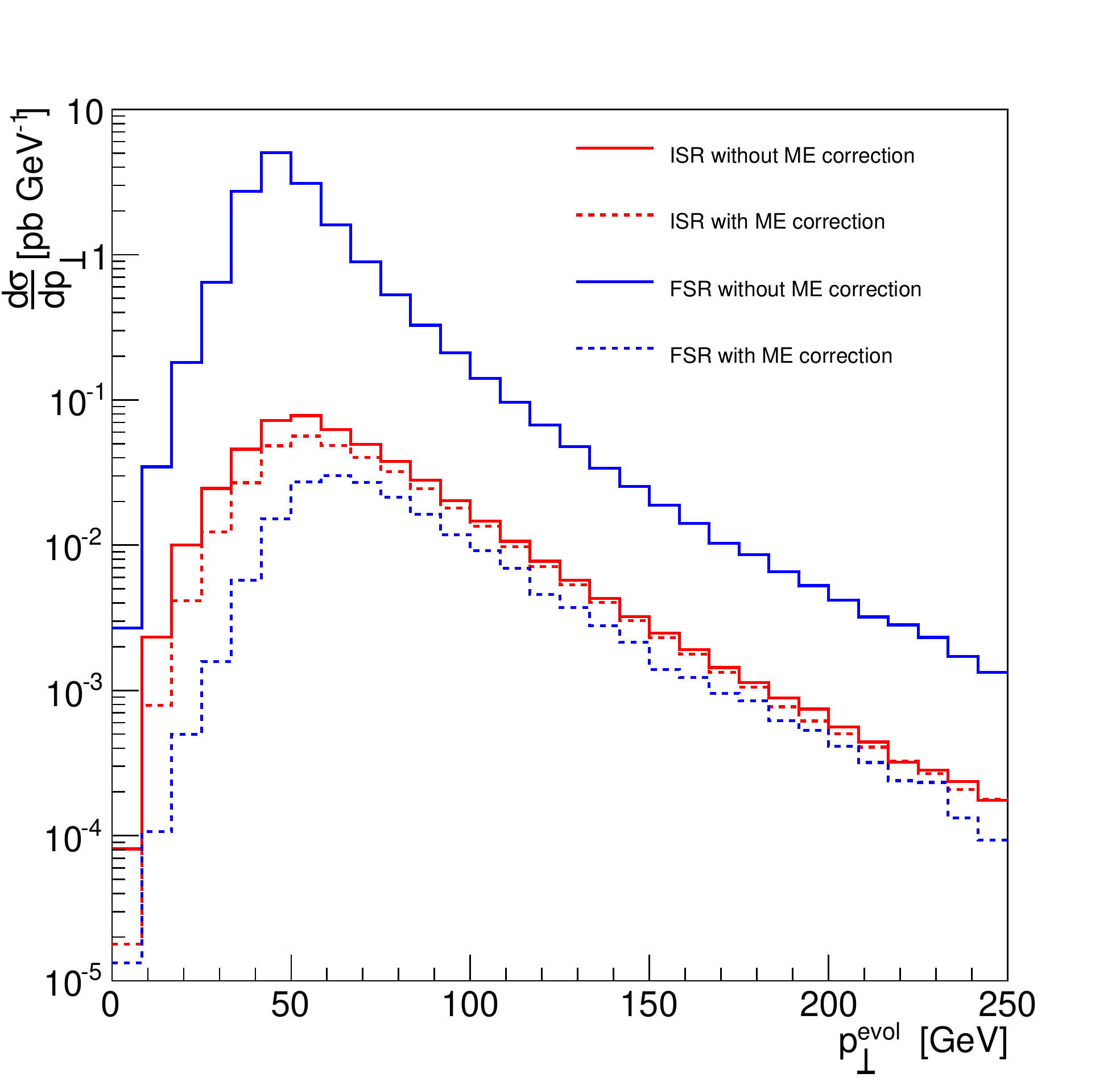}
  \includegraphics[width=0.49\textwidth]{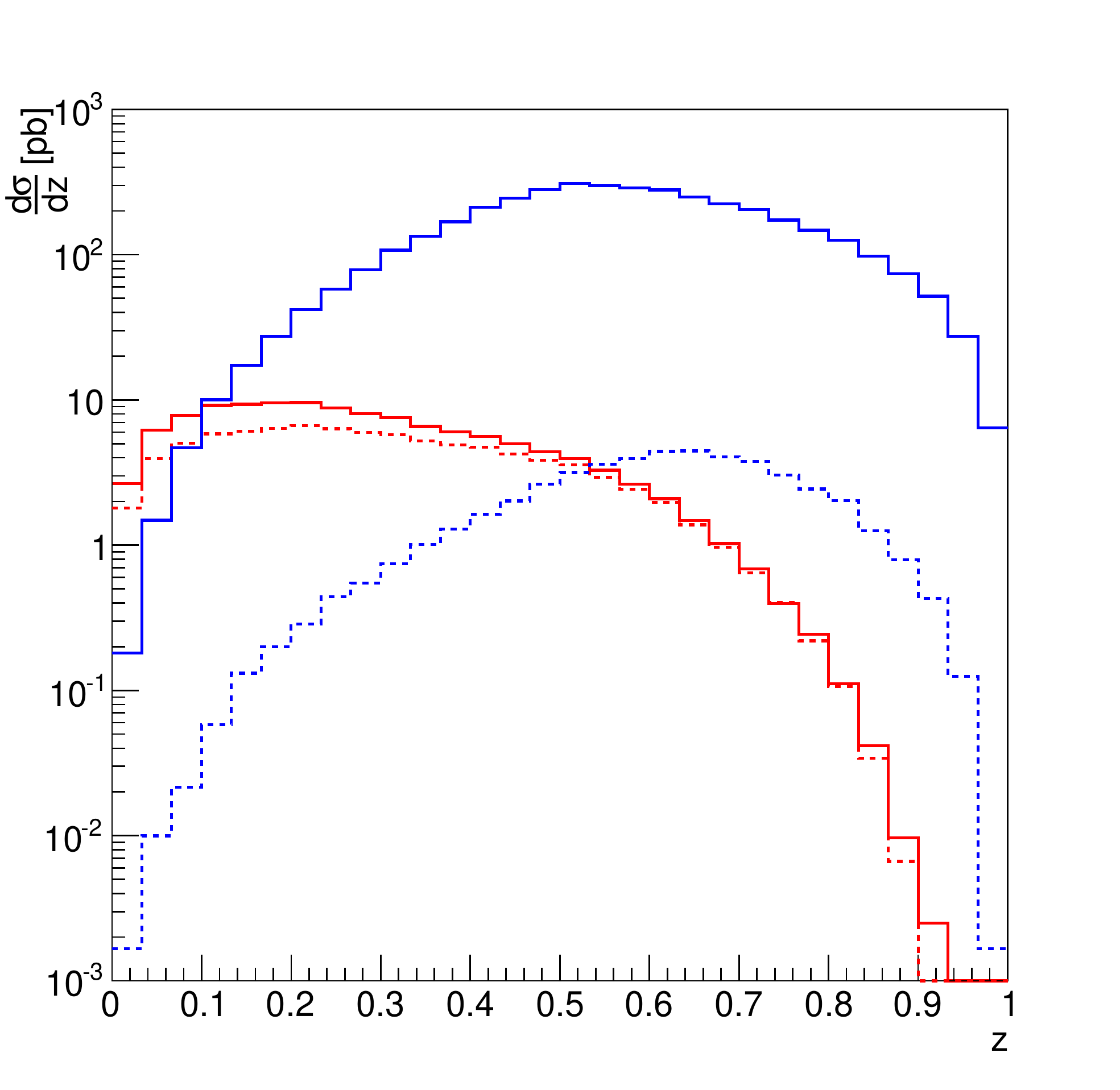}
  (a) \hspace{7.5cm} (b)
  \caption{The differential cross section as a function of (a) $\pTe$ 
    and (b) $z$ for weak boson emission in $s$-channel processes. The
    differential cross sections are shown both with and without including 
    the ME corrections and are separated into ISR and FSR. The center of 
    mass energy was 7~TeV and the minimum $\PT{hard}$ was set to 50~GeV.
    \label{fig:PSvsME}}
  
\end{figure}
 
Results for the $s$-channel process, as a function of the evolution
variable, can be seen in Fig.~\ref{fig:PSvsME}. The FSR results are
obtained with $N = 8$ in eq.~(\ref{eq:PqqZFSR}), and so the rather
crude overestimate of the ME expression is not unexpected. The ISR uses an
overestimate specifically designed for the weak shower eq.~(\ref{eq:PSISR}),
which does a better job at imitating the behaviour of the ME. The
difference between the two curves is largest for small $\pTe$, whereas
for larger momenta the agreement improves. This is expected since the
mass of the weak bosons is more important in the low-$\pTe$
region. The reason that the PS without any correction does not diverge
for  $\pTe \to 0$ is the purely kinematic restriction from the
emission of a heavy boson. Around the PS peak the ratio between the
uncorrected and the corrected number of events goes above 100. The
generation of weak emissions therefore is rather inefficient, leaving
room for improvements. But it should be remembered that the QCD shower
part produces more trial emissions than the weak shower one does, and
that therefore the overall simulation time should not be
significantly affected by the weak shower. Similar results but as a function
of the energy-sharing variable, $z$, can also be seen in Fig.~\ref{fig:PSvsME}. 
The FSR overestimate has the same structure as the ME and only an overall 
factor differs between the two. The ISR overestimate gets slightly worse for 
high and low values of $z$.

\begin{figure}[tbp]
  \centering
  \includegraphics[width=0.49\textwidth]{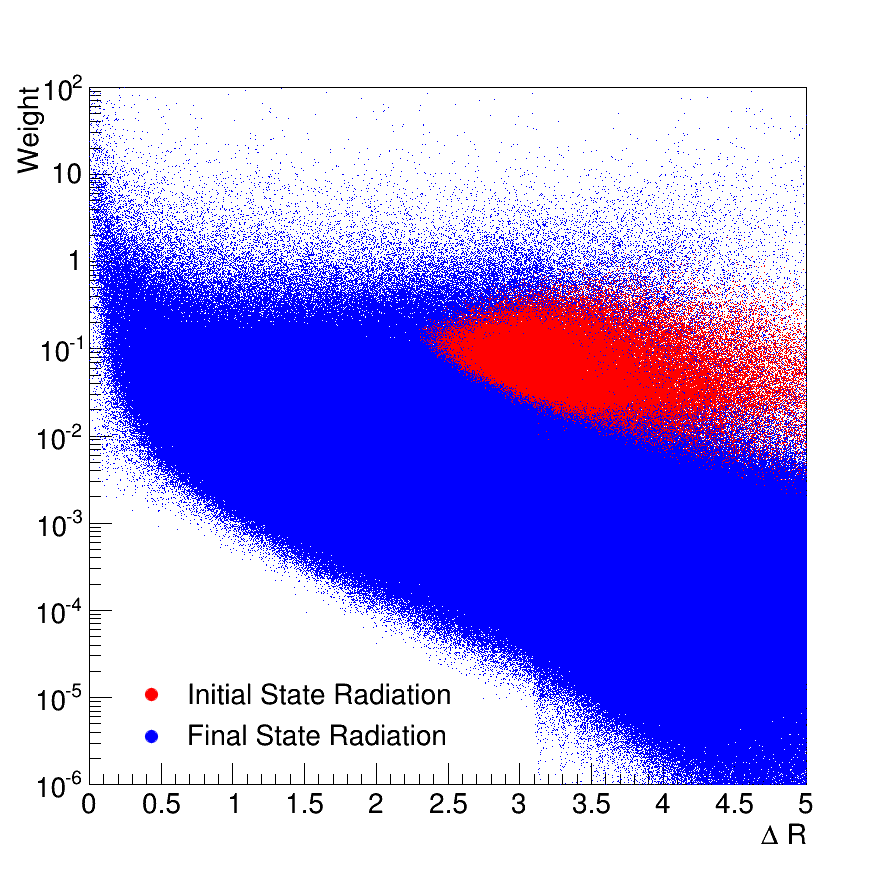}
  \includegraphics[width=0.49\textwidth]{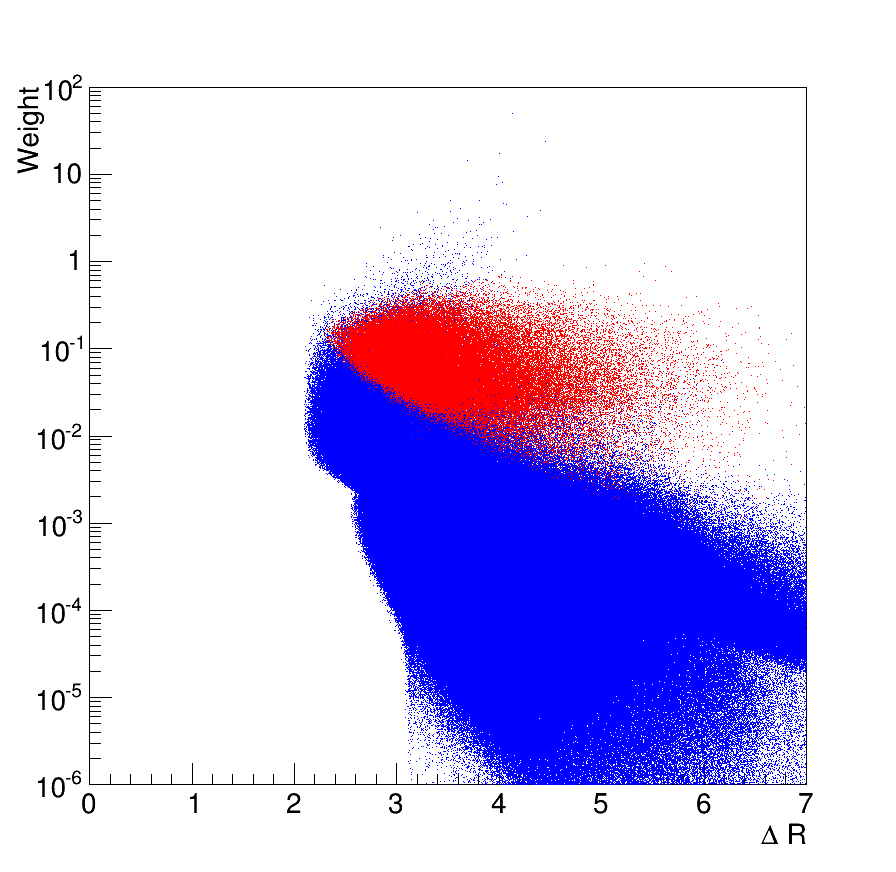}
  (a) \hspace{7.5cm} (b)
  \caption{Scatter plots showing the weight distributions as a
    function of $\Delta R$ between the final state quark and the final 
    state gluon for the process $\u \g \to \u \g \Z$. In (a)
    all trial emissions are included, whereas the no-double-counting cuts have
    been imposed in (b). The ISR points masks part of the FSR ones.
    The starting minimum $\pT$ of the hard
    process was set to 50~GeV and only the weak shower was enabled. $R = 0.6$
    was used in the clustering step for (b).
    \label{fig:weights}}
\end{figure}

For the $t$-channel processes the PS is not guaranteed to be an
overestimate of the ME. Indeed, for all the processes there are
emissions with weights above unity, and significantly so. This
indicates a divergence in the ME that is not covered in the PS. It
turns out that the problematic type of events contain a low-$\pT$
parton in the final state, quark or gluon, or two almost collinear
partons. This can be seen in  Fig.~\ref{fig:weights} for the $\u \g
\to \u \g \Z$ process, where the weight becomes high as the quark becomes
collinear with the gluon. These types of
events were discussed in the double-counting section: in a PS approach
they should be produced by a Drell--Yan weak boson production followed
by QCD emissions, and not by emission of weak bosons within the
PS. Once the doublecounting cuts are introduced,
Fig.~\ref{fig:weights}, the weights are much better behaved than
before. The hard cutoff at $\Delta R = 2\pi/3$ is due to momentum
conservation in the three-particle final 
state. Some very few phase-space points remain with weights above 
unity. Events in such points are always accepted,
but only with the standard unit weight, so the PS  produces too few
events in these regions. Given the tiny fraction involved, this
effect should be small in most phenomenological studies. Similar to what was
seen for the $s$-channel overestimate, the ISR overestimates behave
nicely. Almost all the ISR weights stay within a band between 0.01 and 1. In
addition to the aforementioned problem with weights above unity, the FSR has
a large bulk of trial emissions with very low weights, making the generation
inefficient. 

\subsection{Transverse momentum distribution of the weak boson}

It is possible to validate the implementation of the PS by comparing
it to ME calculations for $2 \to 3$ processes, and to this end we have
used CalcHEP to generate events according to various MEs. Strictly
speaking this only ensures that the first emission is correct and
does not reveal anything about how the PS describes later emissions.

In order for the comparison to be meaningful, the Sudakov factors have
to be removed from the PS. This can be achieved by a veto on each
emission after statistics on it has been collected.  The evolution
thus continues downwards from the vetoed scale as if  there had been
no branching, which cancels the Sudakov. To match the choices made for
the ME calculations, the factorization and renormalization scales are 
held fixed at the $\Z^0$ mass throughout the shower evolution. 
In addition the starting scale for the  emissions was set to  
$\sqrt{\s}$ in order to fill up the full  phase space.

\begin{figure}[tbp]
  \centering
  \includegraphics[width=0.49\textwidth]{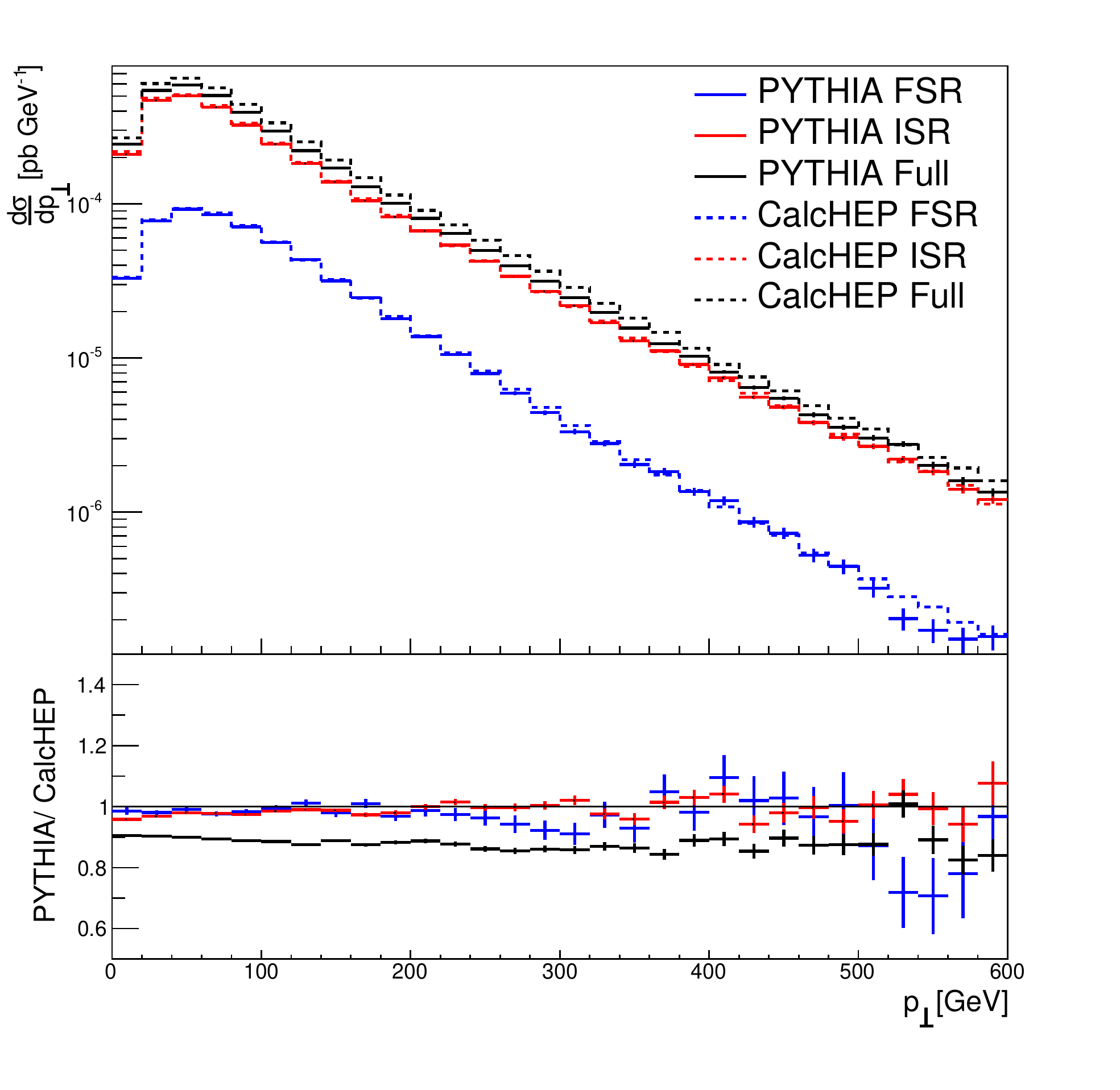}
  \includegraphics[width=0.49\textwidth]{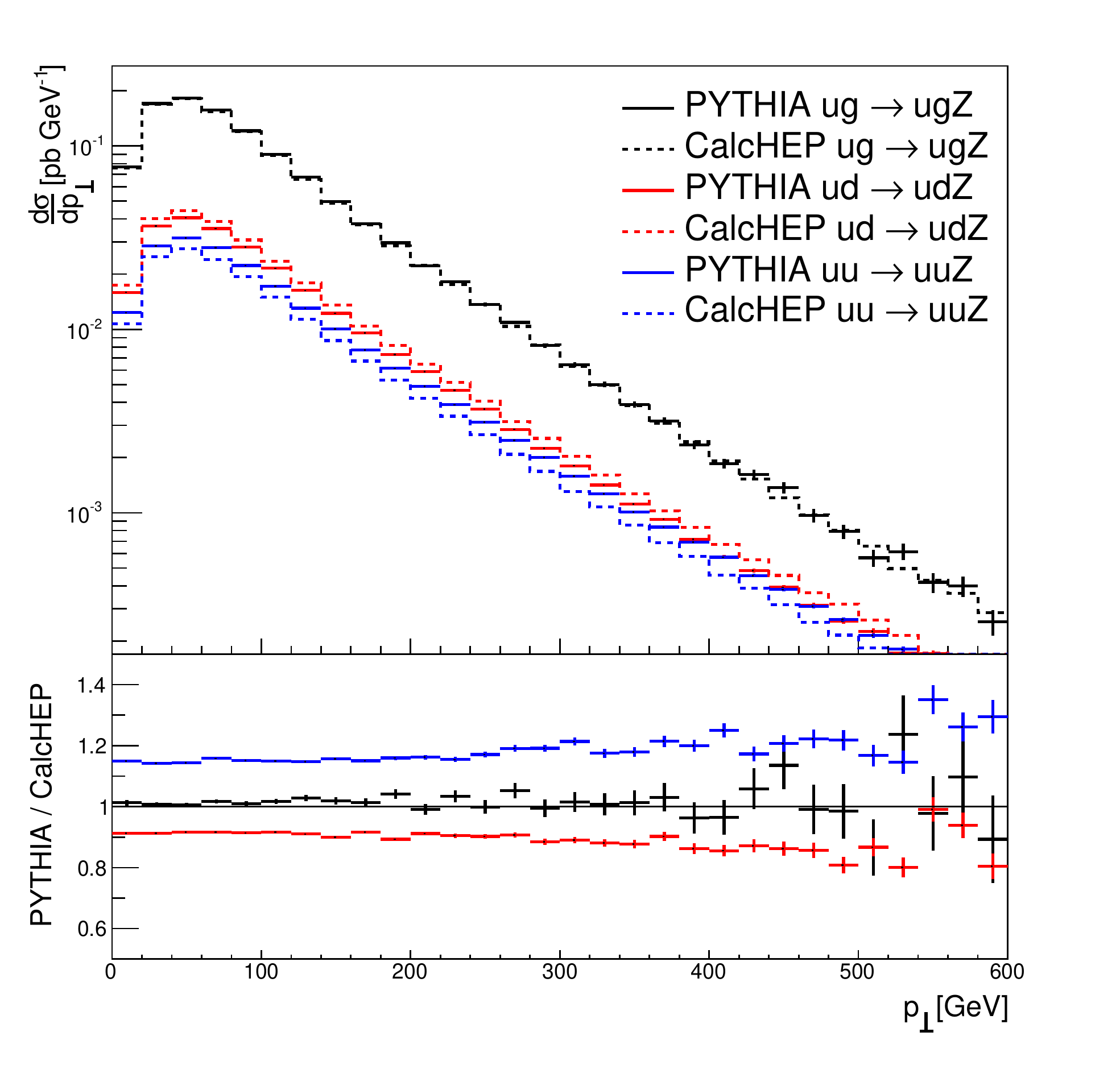}
 (a) \hspace{7.5cm} (b)
  \caption{Comparison between CalcHEP and \textsc{Pythia}~8 results for
  representative (a) $s$-channel $\d \bar{\d} \to \u \bar{\u} \Z$ and
  (b) $t$-channel processes. The center of mass energy was
  8~TeV and the following phase-space cuts were applied
  to avoid divergent regions: $p_{\perp\u} > 100$~GeV, 
  $p_{\perp\g} > 100$~GeV, and $M_{\u\g} > 150$~GeV.  
  \label{fig:pTValidations}}
\end{figure}

For comparisons in pure $s$-channel processes, an ISR and an FSR part
can be read off from the full ME, using the coupling structure. 
It is therefore possible to compare these parts individually
with their respective PS type. Since the PS is already corrected 
to these MEs, a perfect match is expected and also observed, see
Fig.~\ref{fig:pTValidations}. For the combined case, this is  no
longer true, since the PS does not include the interference effects
present in the full ME answer. This difference amounts to the order of 
10\% in the chosen phase-space region, with similar numbers in several 
other phase-space regions that have been tested. The PS is only expected 
to be an approximation to the full MEs, and therefore the discrepancy 
just shows with which accuracy the PS works. 

Since the $t$-channel processes do not admit a natural split between 
ISR and FSR, only the combined results are relevant. The  
$\u \g \to \u \g \Z$, $\u \d \to \u \d \Z$ and $\u \u \to \u \u \Z$ 
comparisons are shown in Fig.~\ref{fig:pTValidations}. For the 
quark-gluon hard process a perfect match is expected and observed, since 
the full ME correction is used. It also shows that, at least in this part
of phase space, the small problem with weights above unity is
negligible (not a surprise given the cuts chosen). The  
$\u \d \to \u \d \Z$ and $\u \u \to \u \u \Z$  cases both
have a discrepancy between the MEs and the PS. In both cases
this comes from the PS ignoring interference between emissions 
of weak bosons from different fermion lines. For the latter case 
there is a further problem, namely that the applied ME correction is that of
$\u \d \to \u \d \Z$ and not $\u \u \to \u \u \Z$.

\begin{figure}[tbp]
  \centering
    \includegraphics[width=0.49\textwidth]{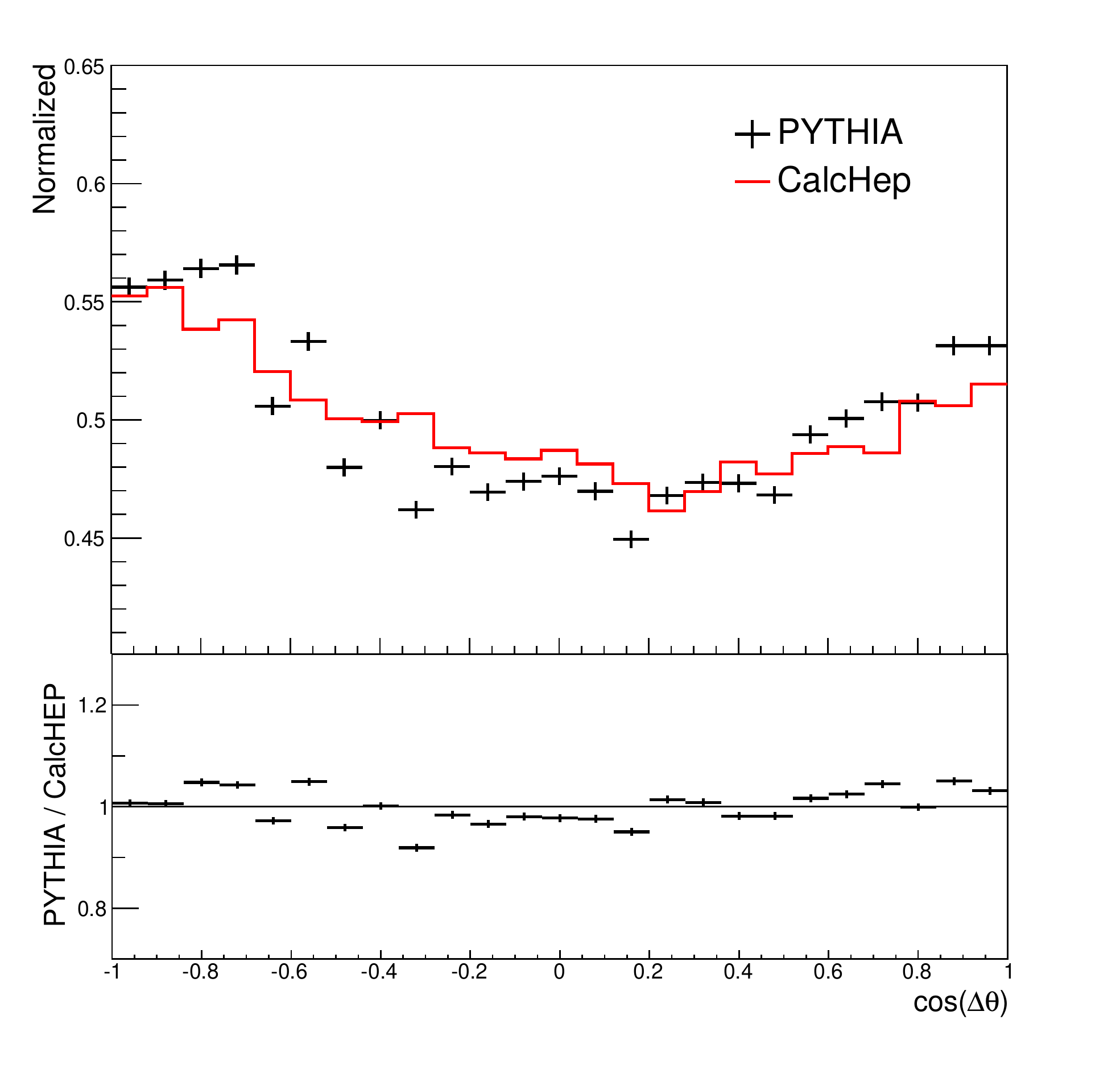}
    \includegraphics[width=0.49\textwidth]{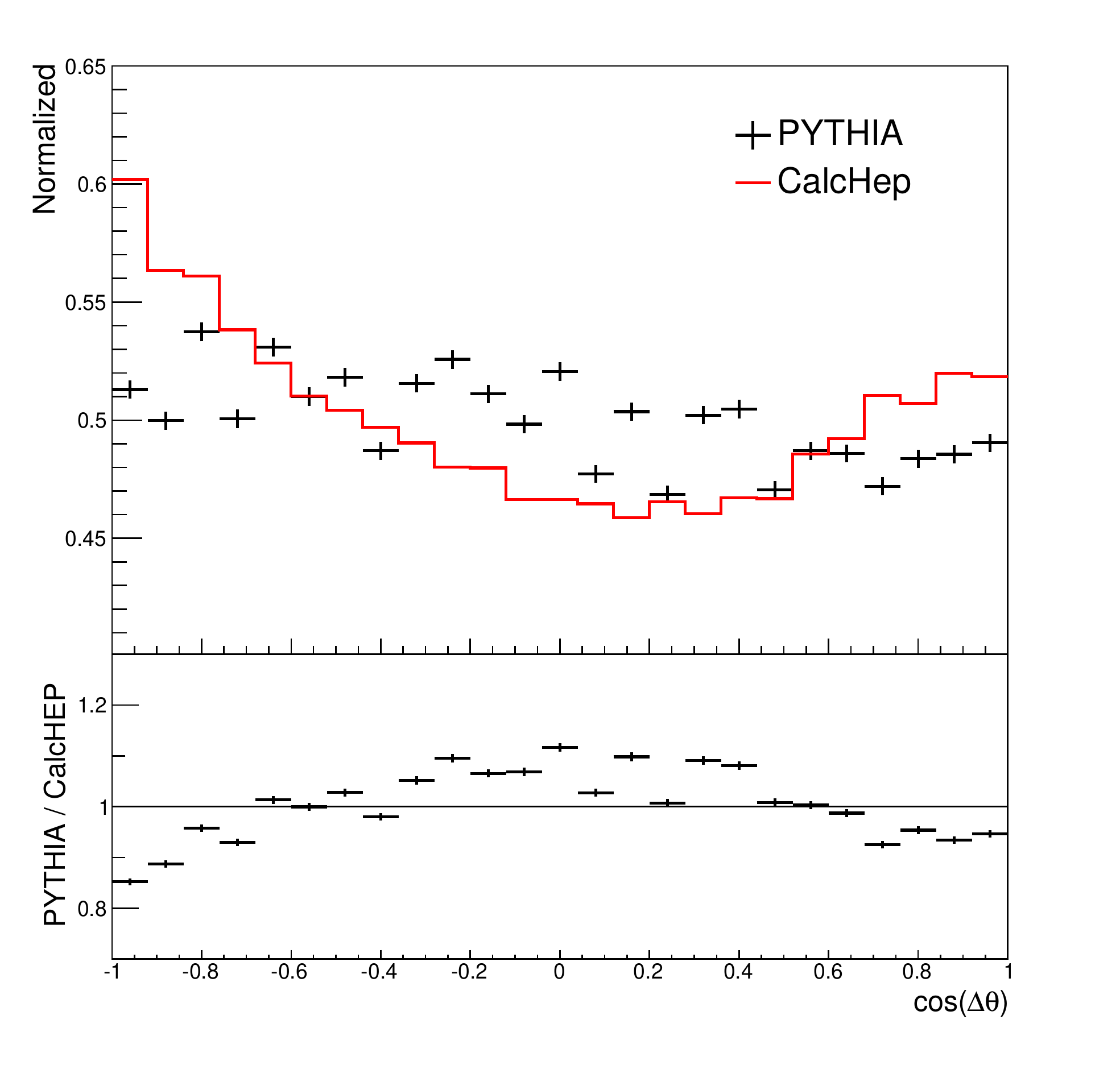}
    (a) \hspace{7.5cm} (b)
    \caption{Angular distributions in the $2 \to 4$ process 
    $\u \g \to \u \g \e^+ \e^-$ between (a) the $\u$
    quark and the electron and (b) the gluon and the electron,
    defined in the $\W$ rest frame. To avoid divergent 
    phase-space regions the same cuts are applied as in 
    Fig.~\ref{fig:pTValidations}.
    \label{fig:decayResults}}
  
\end{figure}

\subsection{Angular distributions of the weak boson decay products}

The angles between the decay products of the weak boson and the other
partons in a $2 \to 4$ process 
(e.g. $\q\qbar \to \q\qbar\Z^0 \to \q\qbar \e^- \e^+$) 
are by construction matched to the angles calculated from the 
correct MEs for $s$-channel processes, separately for ISR and FSR. 
The same $s$-channel-derived corrections are used also for $t$-channel 
processes. To check the validity of this approach, the angular 
distributions between the partons and the leptons have been studied 
for the $\u \g \to \u \g \e^+ \e^-$ case, see Fig.~\ref{fig:decayResults}. 
The angle is defined in the rest frame of the decaying weak boson. 
Since only the shape is relevant, all the curves are normalized to unit 
area. The angle between the quark and the electron is well described
by this ME correction, whereas the same can not be said for the angle
between the electron and the gluon: the PS prediction is almost flat, 
while the ME has peaks around the collinear and anti-collinear
regions. Note the suppressed zero on the $y$ axis of the figure,
however, such that both distributions stay within $\pm$20\%
of being flat. 

The discrepancy could influence relevant observables, 
one obvious candidate being the isolation of leptons: if the
the gluon and  the electron are less likely to be collinear, 
there is a higher probability for the electron to be isolated. 
It should be noted that these calculations are in the rest frame 
of the decaying weak boson, and that the decay products of
a boosted $\Z^0$ will tend to be collimated with each other
and with the emitting parton, away from the $\g$ direction.

\subsection{$\W$ emission in QED hard processes}

So far all the validation tests shown have been for the emission of $\Z^0$ 
bosons. For QCD hard processes and the emission of a single weak boson 
these results translate directly to the $\W^{\pm}$ cases. This does not 
apply for a hard QED process, e.g.\  
$\q\qbar \to \gamma^* \to \e^+ e^-$. Here the emission of a $\Z^0$ can  
be split into ISR and FSR parts, as before, whereas a $\W^{\pm}$ 
additionally can couple to the $\gamma^*$. Then an attempted split
into ISR and FSR becomes gauge dependent, and can individually become 
negative for certain regions of phase space. 

To study this phenomenon, consider the simplest possible $s$-channel 
FSR case, $\gamma^* \to \u \dbar \W^-$, and compare it with the three
related $\gamma^* \to \u \ubar \Z^0$, $\g^* \to \u \dbar \W^-$ and
$\g^* \to \u \ubar \Z^0$ ones. In each case two possible Feynman 
diagrams for the emission of a $\W$ off a quark are included and, 
in Feynman gauge, the squared MEs take the common form 
\begin{eqnarray} 
|M|^2 & = & c \, (\alpha_1 A(x_1,x_2) + \alpha_2 A(x_2,x_1)
+ 2\beta B(x_1,x_2)) ~, \label{eq:ME} \\
A(x_i,x_j) & = & \frac{(1-x_1)(1-x_2)-r^2}{(1-x_i)^2} ~, \\ 
B(x_1,x_2) & = & \frac{x_1 + x_2 - 1 + r^2(x_1+x_2) + r^4}{(1-x_1)(1-x_2)} ~,
\end{eqnarray} 
with the same definitions as before for $x_1$, $x_2$ and $r$. 
The $c$, $\alpha_1$, $\alpha_2$ and $\beta$ are coefficients that depend 
on the specific process. For the three reference processes the 
coefficients are $\alpha_1 = \alpha_2 = \beta = 1$, suitably normalized,
cf.\ eq.~(\ref{eq:MEFSR}).

\begin{figure}[tbp]
  \centering
    \includegraphics[width=0.49\textwidth]{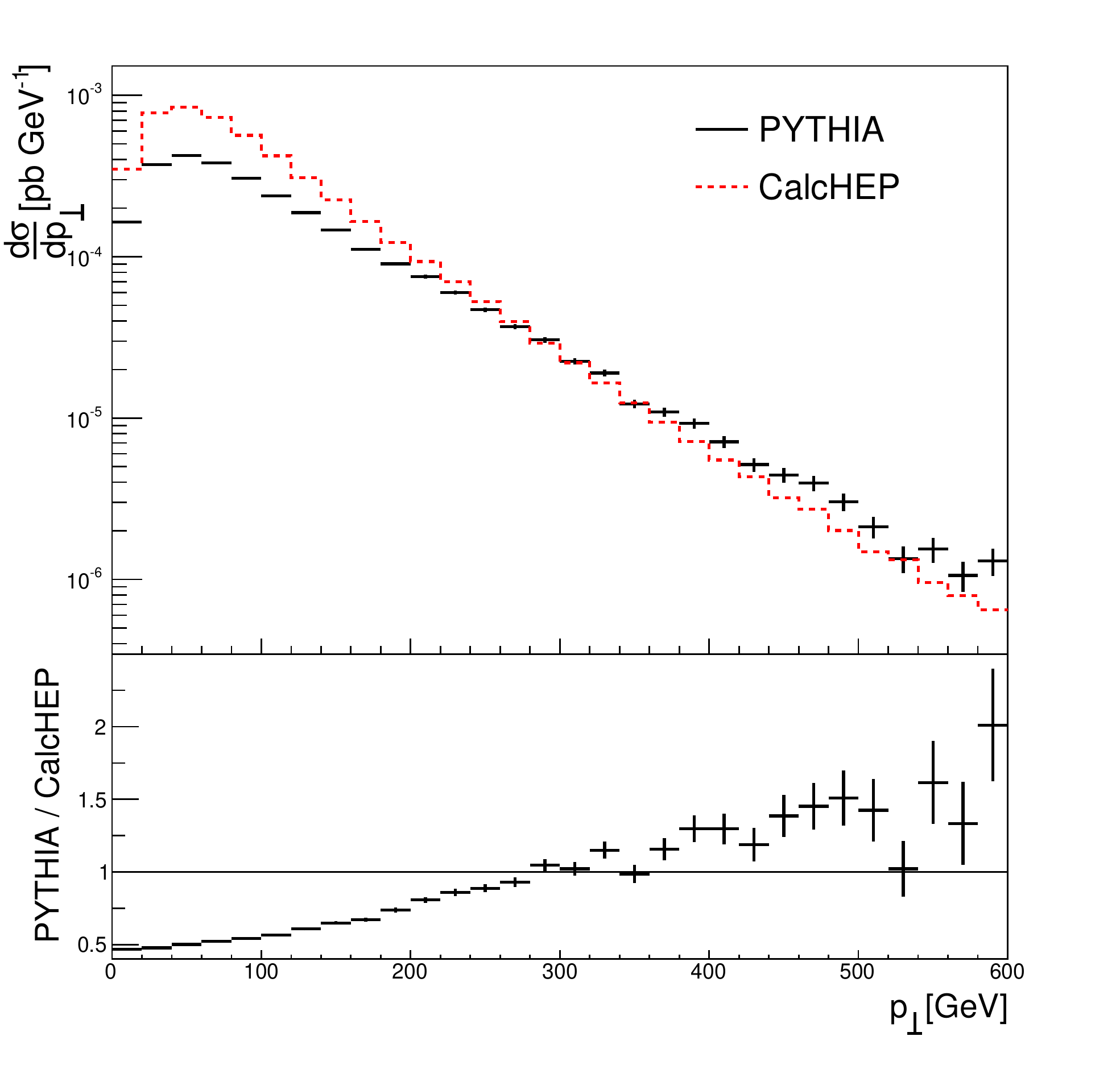}
       \caption{Comparison between the $2 \to 3$ ME calculation and the
         prediction from the PS for the $\u \bar{\u} \to \u \bar{\d} \W^-$
         process, including only electroweak diagrams. Similar cuts as in
         Fig.~\ref{fig:pTValidations} were applied.  
      \label{fig:wEmission}}
\end{figure}

For the $\gamma^* \to \u \dbar \W^-$ process, however, the coefficients 
become $\alpha_1 = e_{\u}^2 = 4/9$, $\alpha_2 = e_{\d}^2 = 1/9$ and
$\beta = e_{\u} e_{\d} = -2/9$. This gives a cross section that is negative
over a large fraction of the phase space.
The reason obviously is that we have neglected a third diagram specific
to this case, involving the triple-gauge-boson vertex 
$\gamma^* \to \W^+ \W^-$, which restores positivity. 

The introduction of a complete electroweak gauge boson shower is beyond 
the scope of this study. For now we therefore handle cases like this
using the same shower couplings and ME corrections as for the cases
with a gluon propagator. 
To estimate the effect of this approximation,
the $2 \to 3$ ME for $\u \ubar \to \u \dbar \W^-$ with all electroweak 
diagrams included, but not QCD ones, was compared to the prediction 
from the PS, see Fig.~\ref{fig:wEmission}. This includes $s$- and 
$t$-channel exchange of $\W^\pm$, $\Z$ and $\gamma$, and is dominated by 
the $t$-channel contributions for the studied regions of phase space.
The comparison looks reasonable for large $\pT$ values, but for small 
values the ME rate is about twice as large as the PS one. 
Such a qualitative agreement should be good enough, given the dominance 
of QCD processes at the LHC. 

\section{Studies of  jet phenomena at LHC energies}

We now turn to studies of how the introduction of a weak
shower changes different observables at the LHC. Three representative
examples have been chosen. Firstly, weak corrections to the exclusive 
di-jet production, and some other generic rate measures. Secondly, 
how likely it is to find a $\WZ$ 
decaying hadronically inside a high-$\pT$ QCD jet. Thirdly, whether it 
is possible to describe the inclusive $\WZ$ + jets cross sections that 
the ordinary PS fails to describe. \textsc{Pythia} version 8.181 was 
used for all the phenomenological studies. The choice of PDF was CTEQ6L
\cite{Pumplin:2002vw}, with a NLO running $\as$.

\begin{figure}[tbp]
  \centering
    \includegraphics[width=0.49\textwidth]{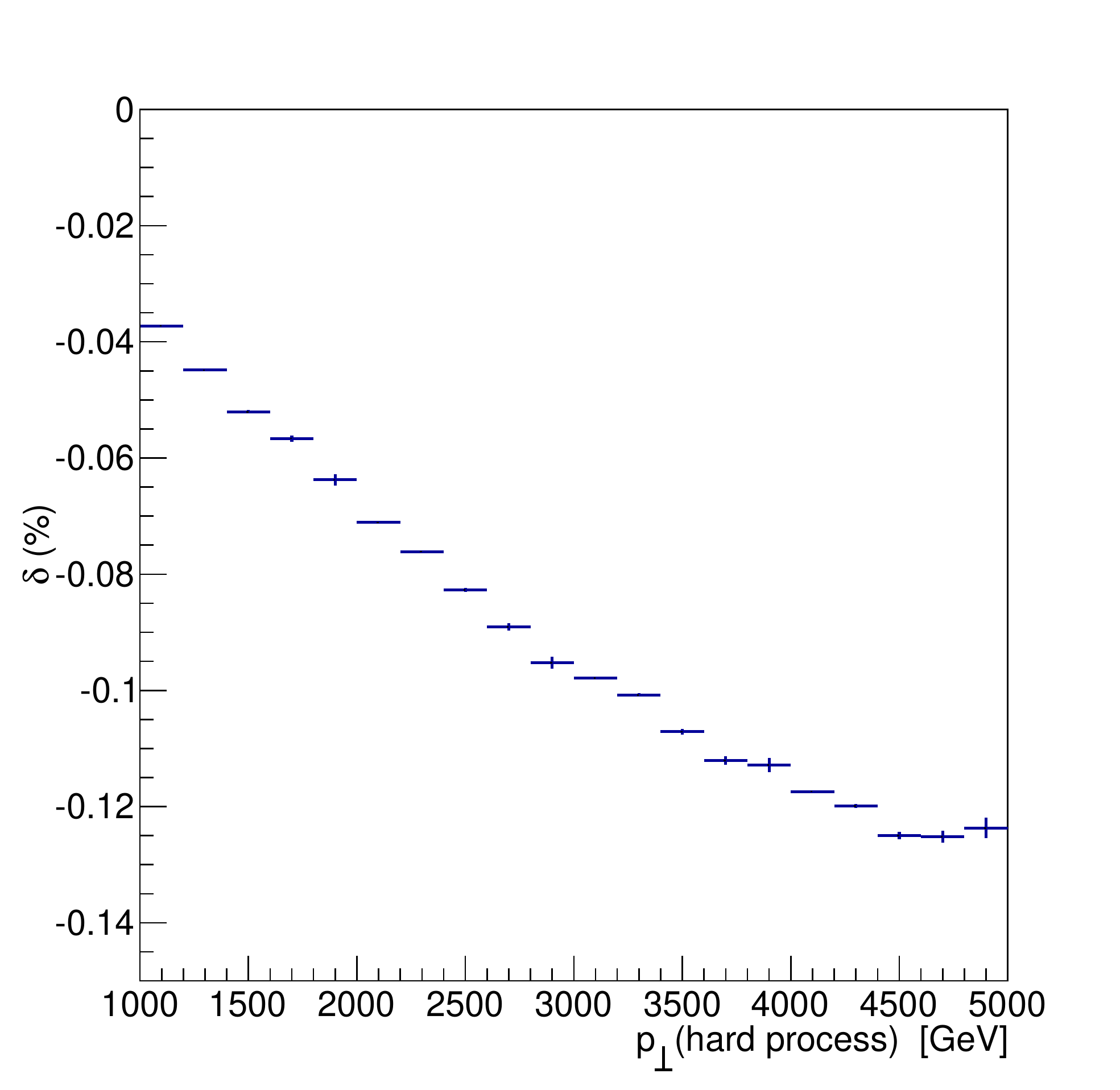}
    \caption{Weak corrections to di-jet production. $\delta$ is 
    calculated by removing all events with a weak boson emission, 
    $\delta = - \sigma(2\to 3) / \sigma(2\to 2)$. 
    \label{fig:diJetWeak}}
\end{figure}

\subsection{Exclusive di-jet studies}

The calculation of Moretti, Nolten and Ross (MNR)
\cite{Moretti:2005ut,Moretti:2006ea} showed large negative
$\mathcal{O}(\as^2\aw)$ corrections to jet production at  hadron
colliders, in the range 10--30\% for jets with $\pT > 1$~TeV.  To put
these numbers in perspective, we simulate $2 \to 2$ QCD hard processes
with only the weak shower turned on.  The weak correction is then
defined by the rate at which at least one  $\WZ$ boson is produced,
Fig.~\ref{fig:diJetWeak}. That rate increases  for larger $\pT$ of the
hard process, partly by the PDF change from  gluon-dominated to
quark-dominated hard processes, partly by the  intrinsic logarithmic
increase of the emission rate. In our calculations the corrections are
only in the range 4--14\%, i.e. less than half of  the corresponding
MNR numbers. The comparison is not fair, however,  since we only study
$\mathcal{O}(\aw)$ corrections to $\mathcal{O}(\as^2)$ hard processes,
whereas MNR additionally includes  $\mathcal{O}(\as)$ corrections to
$\mathcal{O}(\as \aw)$ hard processes.

There is another difference between MNR and our PS, in that MNR 
includes Bloch-Nordsieck violation effects, whereas the PS approach 
is based on exact balance between real and virtual corrections to 
the di-jet production. But, as mentioned earlier, the BN violations 
are expected to be small at the  LHC, and therefore the comparison 
should still be sensible.

\begin{figure}[tbp]
  \centering
  \includegraphics[width=0.49\textwidth]{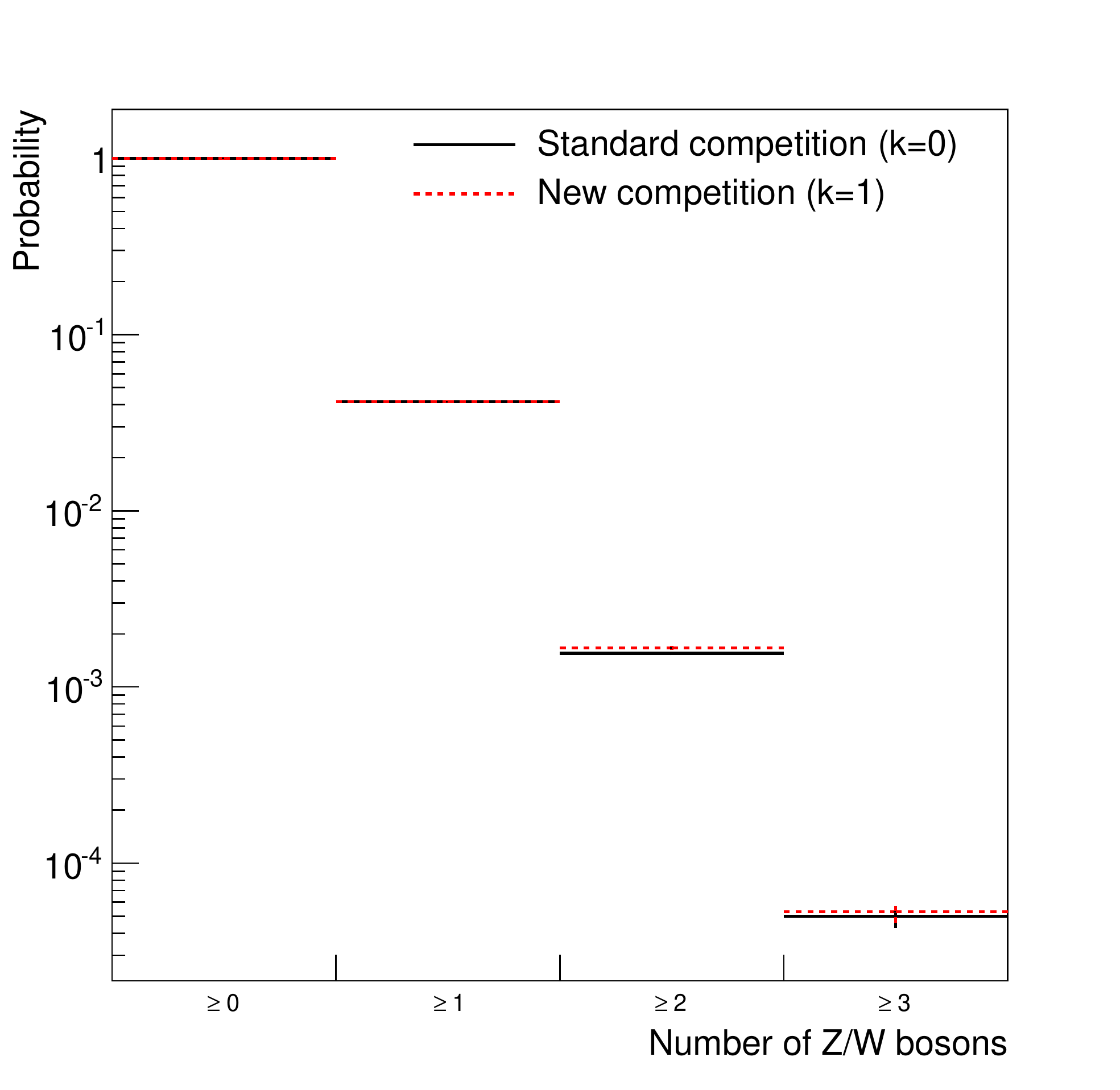}
  \includegraphics[width=0.49\textwidth]{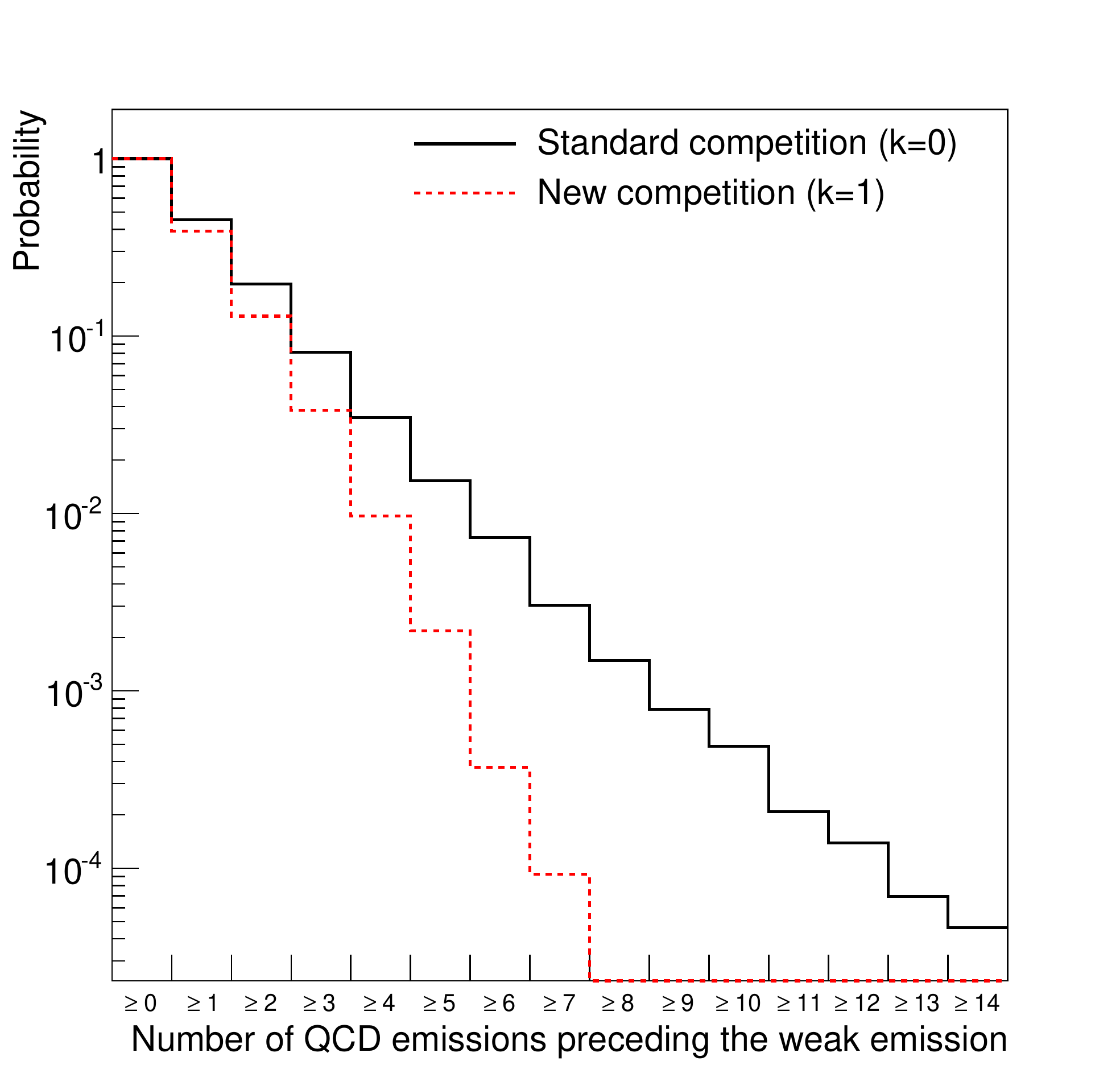}
  (a) \hspace{7.5cm} (b)
  \caption{Probability (a) for multiple emissions of weak bosons and (b) for
    the number of QCD emission preceding the weak emission. The
    center-of-mass energy was set to 14~TeV and the hard process $\pT$ was
    above 1~TeV.  
    \label{fig:QCDvsWeak}
  }
\end{figure}

\subsection{Resummation and competition between the QCD shower and 
the weak shower}

With the availability of a weak shower it is possible to study the
effect of multiple weak boson emissions. This probability is largest 
for high-$\pT$ jets, but even for $2 \to 2$ QCD processes with $\pT > 1$~TeV the
probability for more than a single weak boson emission is found to be 
about one per mille at LHC energies, Fig.~\ref{fig:QCDvsWeak}. For most (if
not all)  
analyses the experimental uncertainty will be significantly above this 
probability, and it is therefore a good approximation to neglect the 
effects coming from other than the first $\WZ$ boson. As already mentioned
the possibility of pure electroweak shower evolution, like 
$\gamma^*/\Z^0 \to \W^+ \W^-$, is not included here.

In a PS it is possible to tell when the weak emission occurs in the 
ordered shower evolution, as opposed to a ME calculation. It is therefore 
of interest to study the competition between QCD and weak emission.
The non-negligible $\WZ$ mass could make for ambiguities, which we explore 
by comparing the normal $\pTe$ ordering with an alternative 
$\pTse + km_{\WZ}^2$ one for weak branchings, with $k=1$ taken as 
the extreme alternative. In Fig.~\ref{fig:QCDvsWeak} we show how 
many QCD branchings occur before the weak one, where a long tail is 
significantly reduced by the larger weak scale choice. The probability
of having at least one preceding QCD emission is of order 40\% in both
extremes, however, underlining that large jets masses are common in
the QCD evolution. Furthermore, the  total number of weak bosons only 
varies by about 2\% between the two choices of evolution scale.
The chances of constraining $k$ from data thus are limited. One 
possibility is to study the energy distributions inside a jet, but the 
problem here is the low experimental rate (to be discussed later). 
Another possibility is weak boson production in association with jets, 
where especially the events with a high jet multiplicity could  be 
influenced by the choice of $k$, a topic that will be studied later 
in this section.

In fixed-order perturbation theory there is no concept of an ordered
evolution, and thus not of how many QCD emissions precede the weak one. 
The shower results here could be an indication of how high orders that
need to be used, in matching and merging schemes, so as to 
describe multijet production in association with a $\WZ$.   
Assuming a ME-to-PS matching at the $\WZ$ emission scale, say, 
Fig.~\ref{fig:QCDvsWeak} suggests that at most 1\% of the $\WZ$ have 
more than five QCD emissions at a higher scale than the $\WZ$ itself
for a $\pT > 1$ TeV jet. Thus, taking into account the two original 
jets, it would be necessary to include up to $\WZ + 7$ jets
so as not to miss relevant multijet topologies down to the 1\% level.  
This is entirely feasible with current technology. Thus the $\WZ$ 
emission in showers does not address otherwise unapproachable 
topologies, but it is likely to address some issues considerably
faster. And jet numbers will rise dramatically for matching scales
below the $\WZ$ one, making showers unavoidable in that region.

\subsection{Substructure of a jet with weak bosons inside the jet} 

If a $\WZ$ is radiated from a high-$\pT$ jet, it is probable that the
$\WZ$ will be boosted along the jet direction, and that the $\WZ$
decay products will fall within a reasonably-sized jet cone. For 
leptonic decays it should still be possible to separate the decay 
products from the other components of the  jet, but for hadronic 
decays the distinction will be more difficult. One possibility is 
to study the jet substructure and jet-jet invariant masses for signs 
of bumps at the $\W$ and $\Z$ masses.

One key challenge is the low $\WZ$ emission rate, around 4\% for
$\pT > 1$~TeV events in total, and not all of that inside jets. 
Another is that the QCD evolution itself produces a rich substructure,
with a high rate in the $\WZ$ mass region. Thus the signal of $\WZ$ 
production inside jets would be small bumps sitting on top of a large 
but smooth QCD background.

\begin{figure}[tbp]
  \centering
  \includegraphics[width=0.49\textwidth]{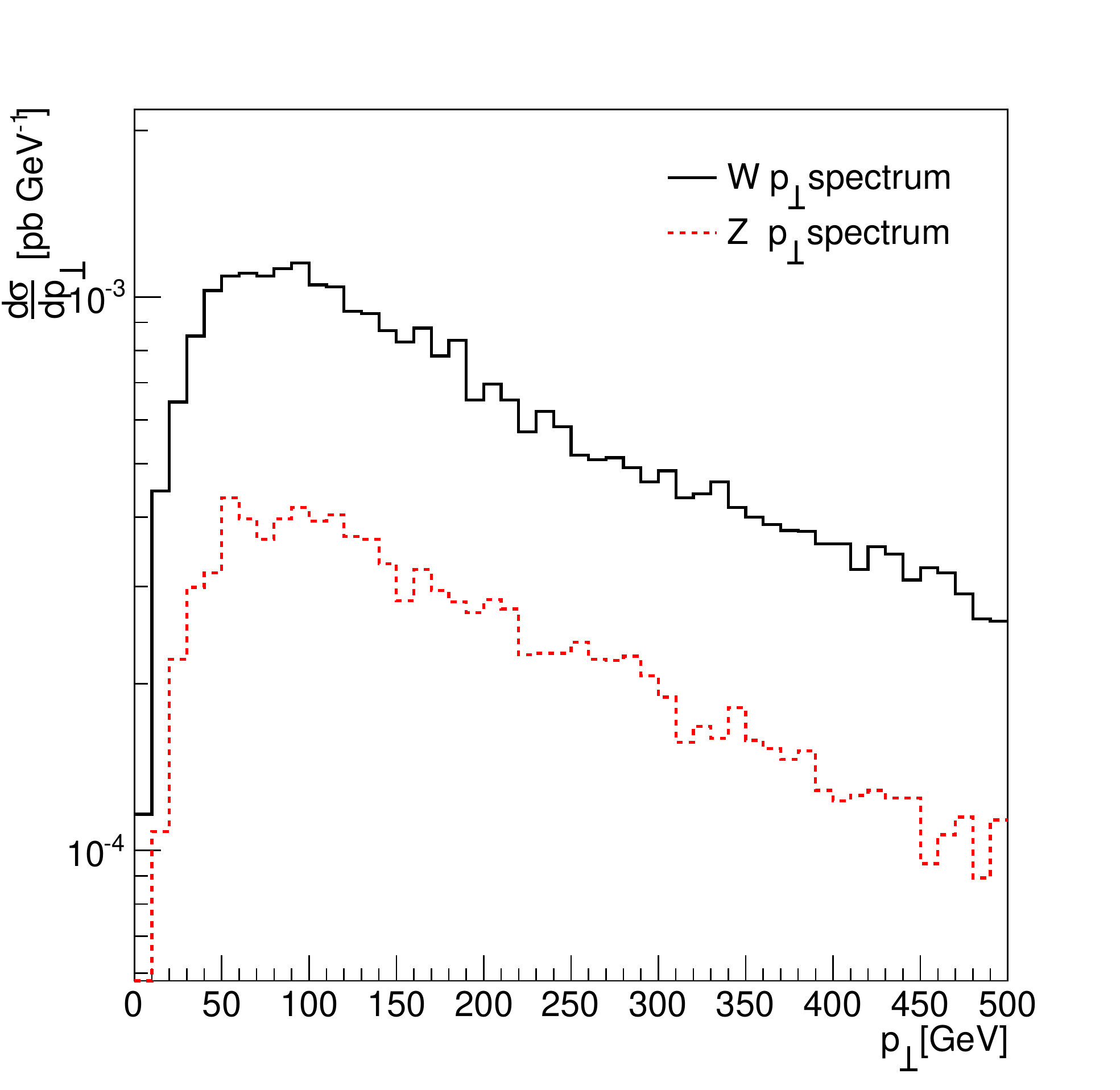}
  \includegraphics[width=0.49\textwidth]{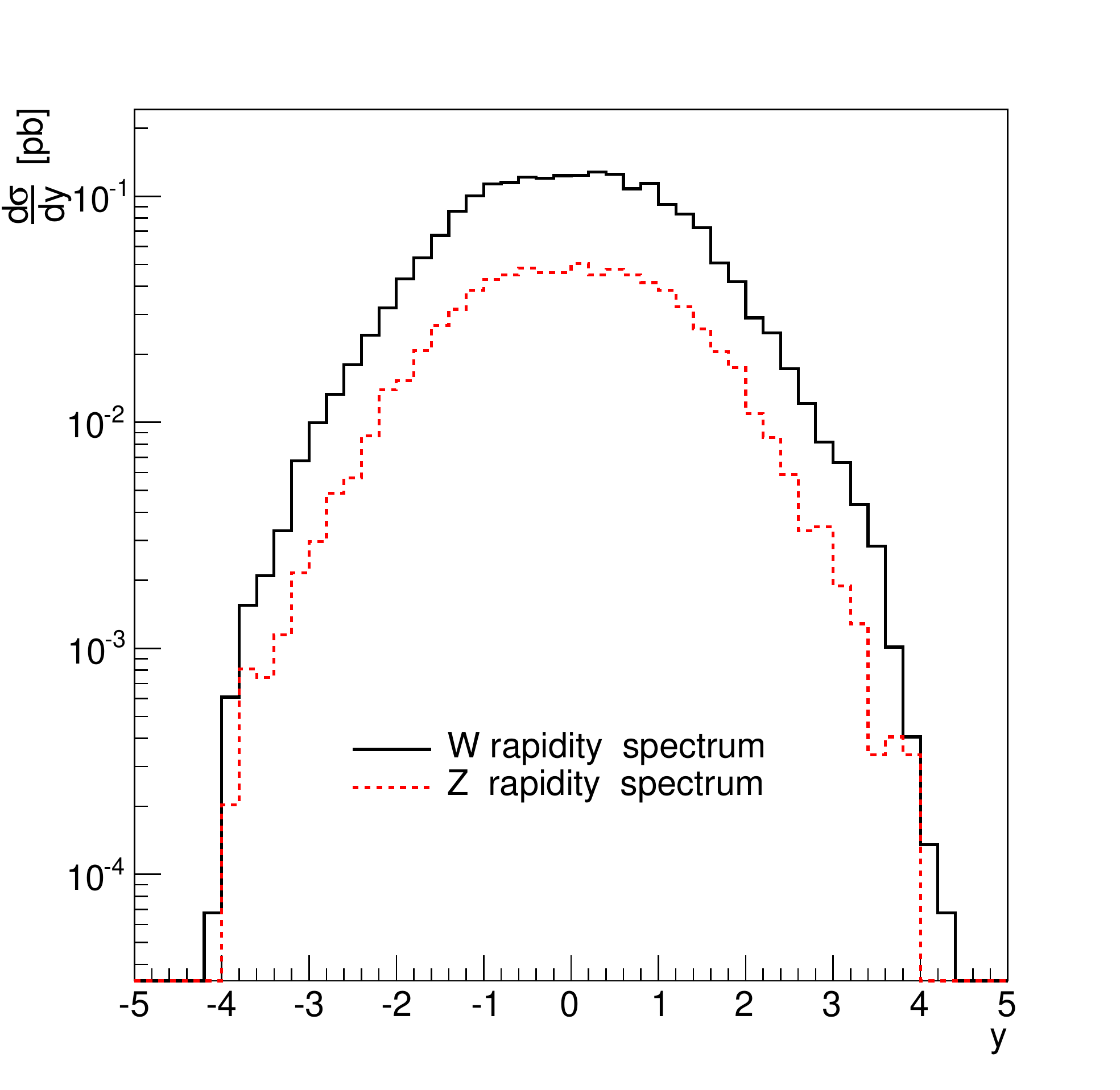}
  (a) \hspace{7.5cm} (b)
  \caption{The (a) $\pT$ and (b) rapidity distributions for $\WZ$ emissions, 
    for $2\to 2$ QCD processes with a transverse momentum above 1 TeV.
    \label{fig:ZWpTrap}
  }
\end{figure}

Before searching for $\WZ$s inside jets, let us take a step back and consider
some of the more basic properties of $\WZ$s produced inside jets,
Fig.~\ref{fig:ZWpTrap}. It is
expected that the $\pT$ distribution for the weak bosons will be significantly
harder than for those produced in Drell-Yan. The Drell-Yan production peaks at
a few GeV, whereas the emissions peak at the mass of the weak bosons,
mainly by simple phase-space effects in the PS.

\begin{figure}[tbp]
  \centering
  \includegraphics[width=0.49\textwidth]{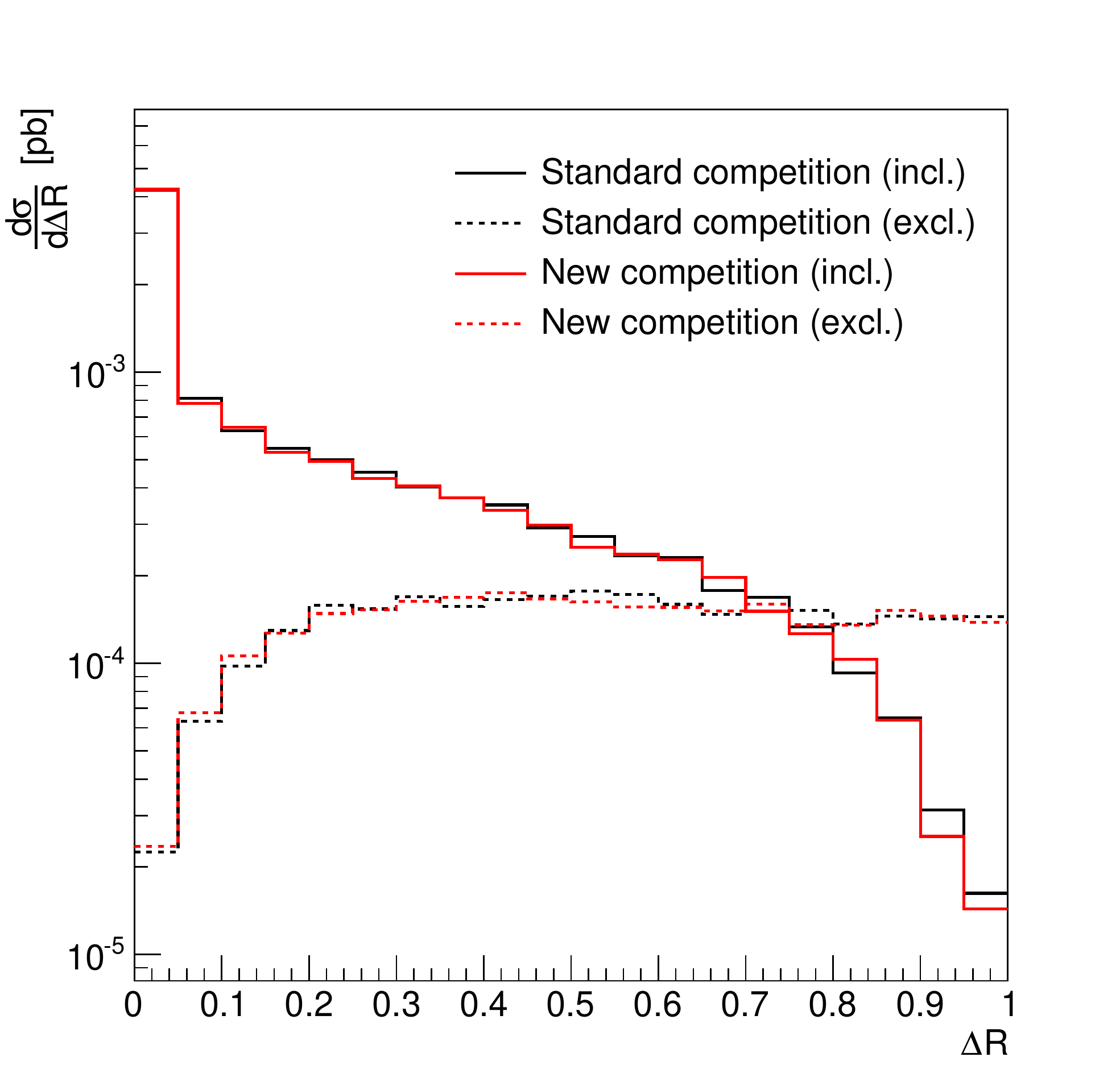}
  \includegraphics[width=0.49\textwidth]{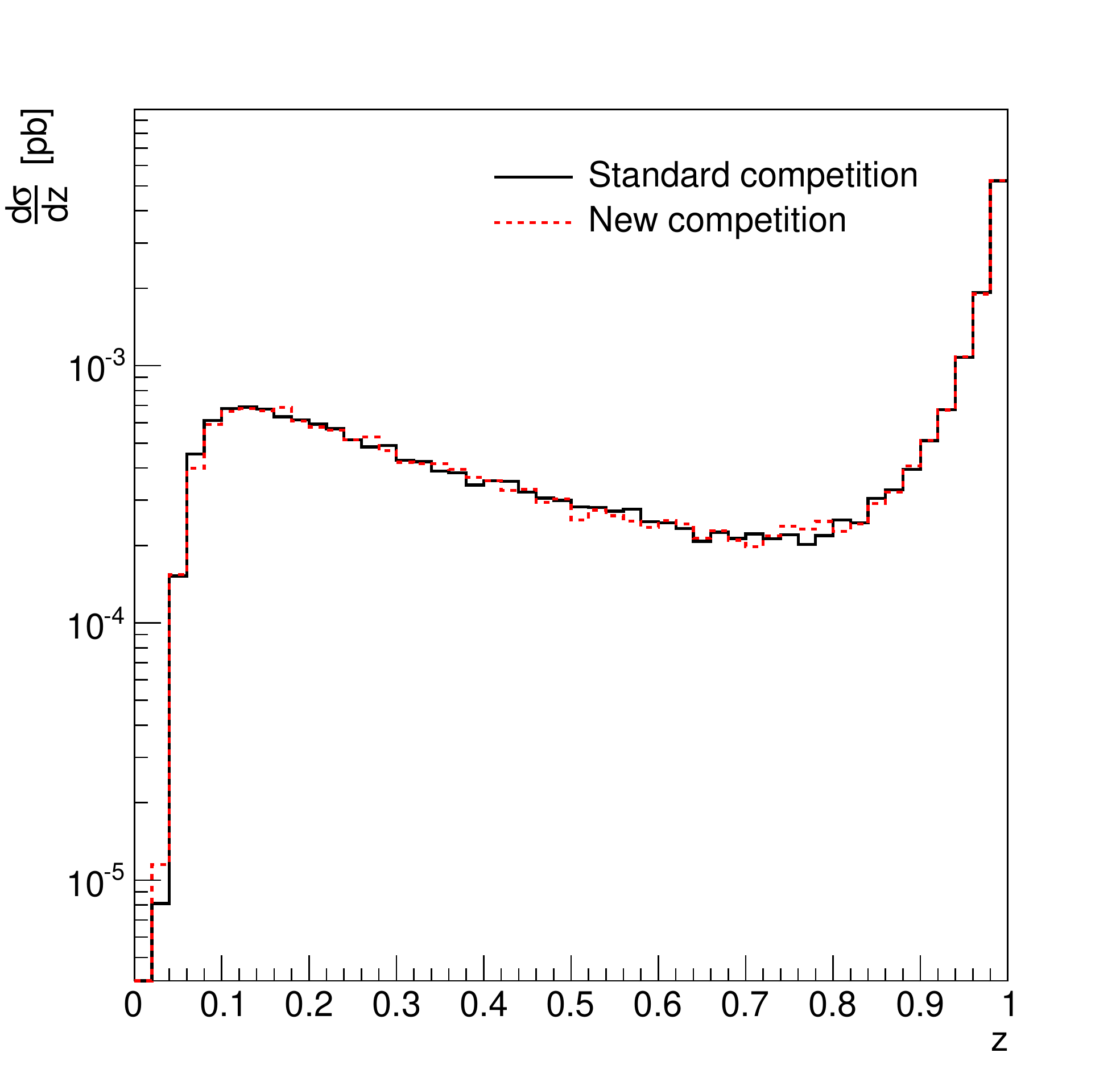}
  (a) \hspace{7.5cm} (b)
  \caption{The (a) $\Delta R$ and (b) energy sharing distributions for radiated
    $\WZ$s. The anti-$k_{\perp}$ jet algorithm has been used with $R = 1$ and
    minimum transverse momentum of 100 GeV. The hard process was $2 \to 2$ QCD 
    processes with transverse momentum above 1 TeV. The $\Delta R$ 
    distributions were calculated both with (inclusive) and without (exclusive)
    including the weak boson in the jet clustering algorithm.
    \label{fig:ZWDeltaR}
  }
\end{figure}

The location of the weak boson within the jet is different from that of the
normal QCD emission. This can be most easily realized by considering the
$\Delta R$ and energy sharing distributions, Fig.~\ref{fig:ZWDeltaR}. 
The energy sharing distribution
for QCD has a peak around zero due to the soft divergence. Weak emissions do
not have soft divergencies, but instead have a hard cut-off due to the mass.
The peaks observed at
respectively $\Delta R = 0$ and $z = 1$ are due to isolated weak bosons
produced from either ISR or large-angle FSR emissions. This is in agreement
with the $\Delta R = 0$ peak disappearing when the weak boson is excluded 
in the jet clustering.

The above distributions are shown for two different competition schemes, the
standard $\pTe$ ordered and the new $\pTse + M_{\WZ}^2$ ordered. The two
schemes give essentially the same results for all the distributions and thus 
none of the observables provide any separation power between the schemes.

\begin{figure}[tbp]
  \centering
  \includegraphics[width=0.49\textwidth]{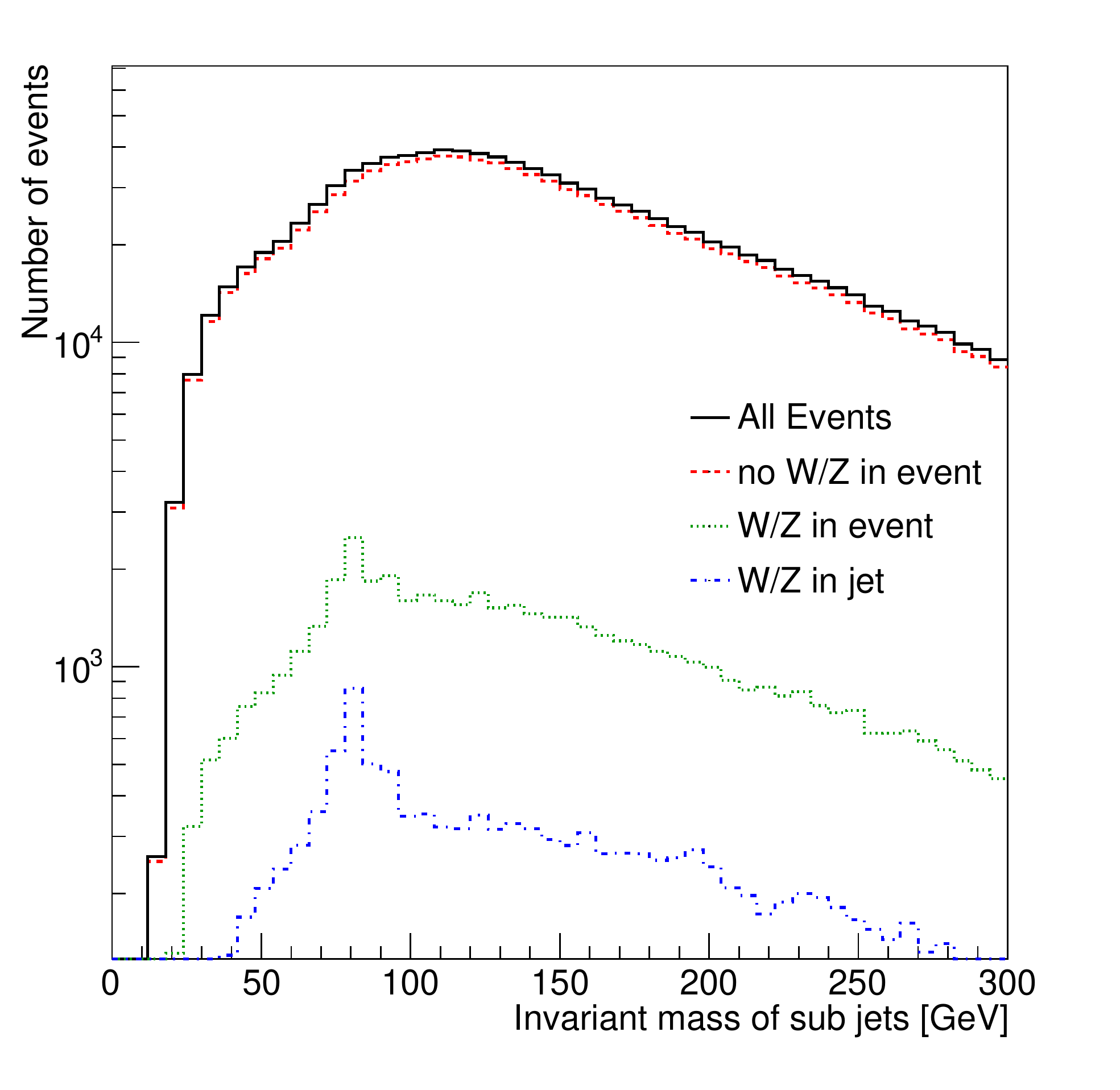}
  \includegraphics[width=0.49\textwidth]{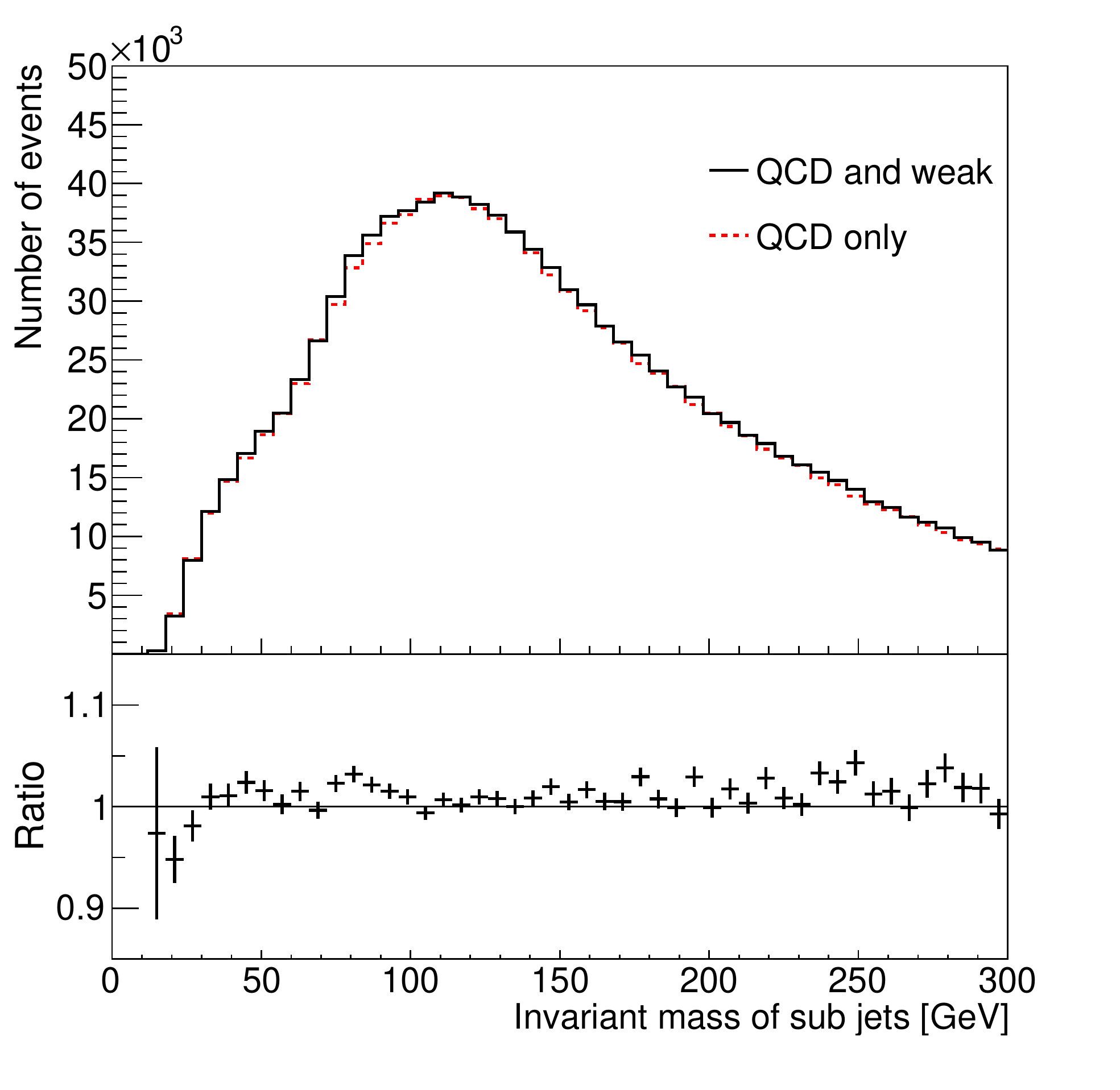}
  (a) \hspace{7.5cm} (b)
  \caption{The invariant mass of all subjet configurations
    found by applying trimming method to a fat jet. In (a) the
    distribution for several different event selections. In (b)
    a comparison between simulations with or without the weak 
    shower included. 
    \label{fig:jetStruct}
  }
\end{figure}

Returning to the hadronically decaying $\WZ$ inside jets, a simple
phenomenological study has been carried out to determine whether is possible
to locate these. First all jets with $\pT > 1$~TeV are found according to the
anti-$\kT$ algorithm with $R = 1$, using FastJet \cite{Cacciari:2011ma}.  
Afterwards these jets are split into subjets by the trimming algorithm, with  
parameters $Z = 0.05$ and $r_{\mathrm{tr}} = 0.2$. The invariant mass of all 
subjet pairs is shown in Fig.~\ref{fig:jetStruct}. Unfortunately no peak 
is visible to the naked eye, due to the large background. As a check, 
if only those events are singled out that contains a $\WZ$ the signal 
stands out, and even more so for those jets that contain a $\WZ$.

As a further step, mass spectra are compared with or without weak
radiation in the shower, Fig.~\ref{fig:jetStruct}, together with the 
ratio. It is possible to see a difference between the two runs, but 
note that only statistical uncertainties have been included (1\,000\,000
events, corresponding to about $77~\mathrm{fb}^{-1}$), and not any 
detector smearing effects.

\subsection{Weak boson production in association with jets}

Inclusive weak boson production in association with jets for a
long time has been known to be poorly described by the \textsc{Pythia}
shower approach alone. The PS can describe the emission of the first 
jet to a decent level of agreement with data, but for additional jets the 
shower predictions fall below the observed data, increasingly so
for more jets. The use of ME corrections for the first emission
is not the reason it works out there: as we have already discussed, 
the uncorrected PS overestimates the $\q\qbar \to \Z\g$ process and 
underestimates the $\q\g \to \Z\q$ one, but by small amounts over most 
of phase space, and in such a way that the sum of the two comes out 
approximately right. Rather we would like to attribute the problem to
the absence of weak emissions in showers, and are now in a position to
check this hypothesis.

The new shower framework has been compared with the $\W$ + jets and 
$\Z$ + jets data from the ATLAS experiment \cite{Aad:2013ysa,Aad:2012en} 
using the Rivet framework \cite{Buckley:2010ar}. The \textsc{Pythia} 
results are obtained as the sum of two components, one ``weak (production)
path'' where the starting topology for shower evolution is a $\WZ$,
and the other a ``QCD (production) path'' where the starting topology 
is a $2 \to 2$ QCD process. The former, being leading order, is well
known to miss out on an overall $K$ factor, which we address by 
normalizing this component to the inclusive $\W$ rate. For the 
$\Z$ production this rate is not quoted, so we instead normalize to 
$\Z + 1$ jet. Empirically there does not seem to be a corresponding 
need for a large $K$ factor in QCD $2 \to 2$, in the context of tunes 
where $\as$ is among the free parameters, and so none has been used here.
Doublecounting between the two paths is avoided as already discussed.

\begin{figure}[tbp]
  \centering
    \includegraphics[width=0.49\textwidth]{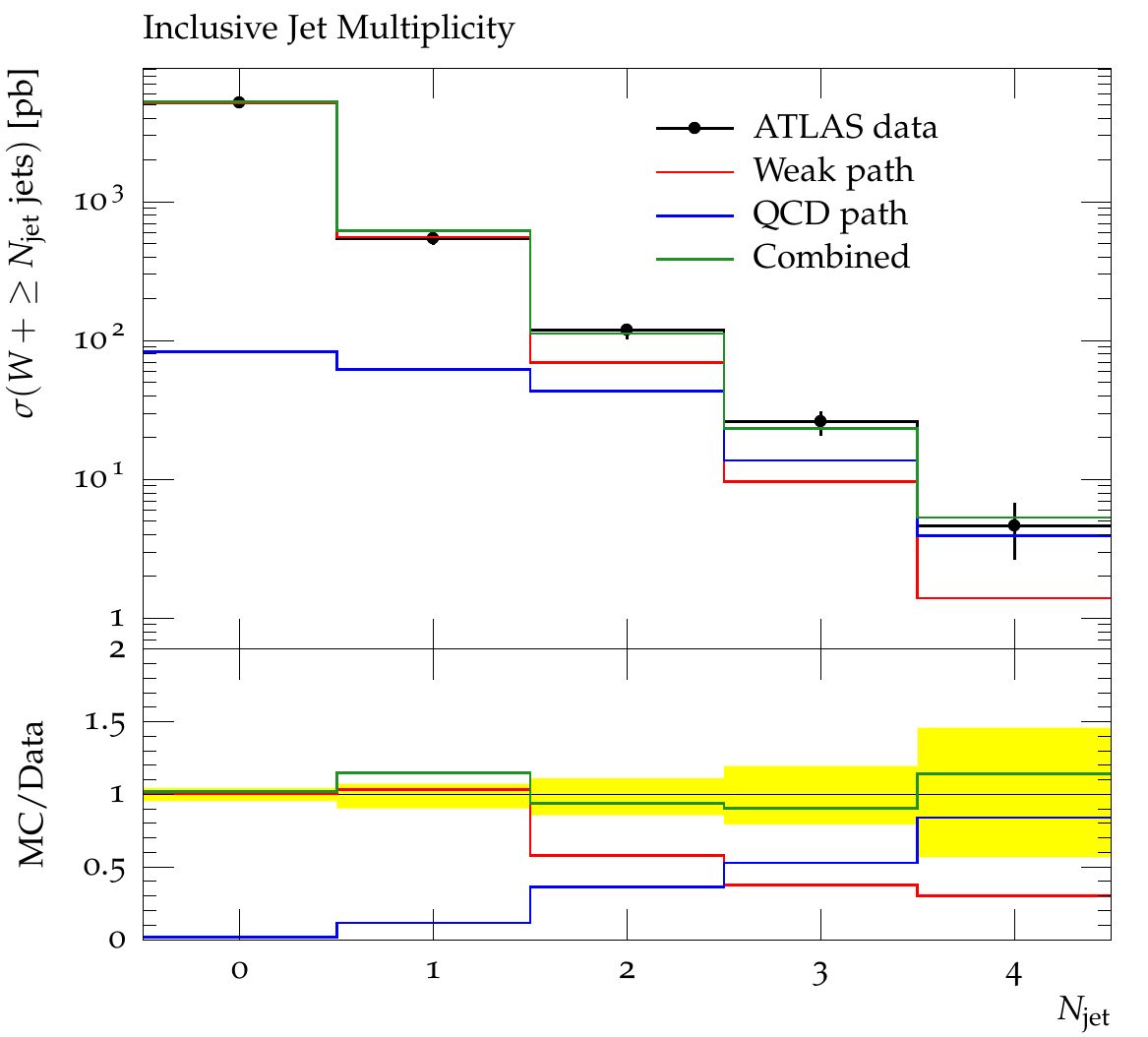}
    \includegraphics[width=0.49\textwidth]{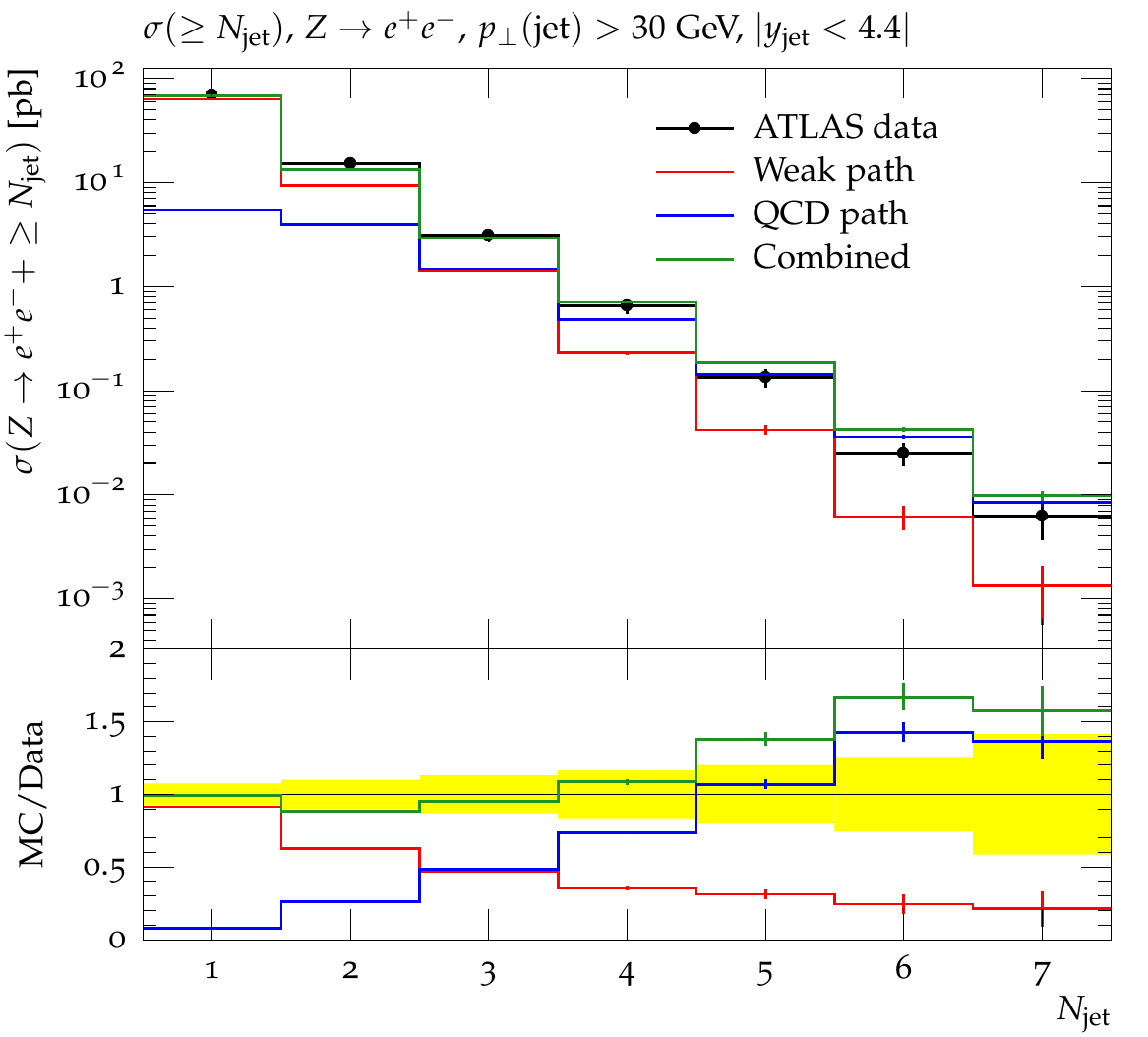}
    (a) \hspace{7.5cm} (b)
    \caption{The inclusive jet multiplicity rates for
      (a) $\W$ + jets and (b) $\Z$ + jets. 
      \label{fig:wJetInclusive}}
   
\end{figure}

The inclusive $\WZ$ cross sections as a function of the number of
associated jets are shown in Fig.~\ref{fig:wJetInclusive}. The weak
production starts to fall off at higher jet multiplicities, as
foretold, where the QCD path  becomes the dominant production
channel. It is clear that the addition of the QCD path, absent in
previous comparisons between data and  \textsc{Pythia}, plays a key
role in achieving a much improved agreement with data. The agreement for the
first 
four jets is very good both for $\W$ and $\Z$, the only slight problem being
the $\W + 1$ jet bin. For higher jet multiplicities the PS start to
overestimate the production. The discrepancy might very well be within
tuneable parameters, for instance a small change in $\as$ will have a large
influence at high jet multiplicities. And given all the approximations
made, the overall agreement might be better than one had initially expected.

\begin{figure}[tbp]
  \centering
    \includegraphics[width=0.49\textwidth]{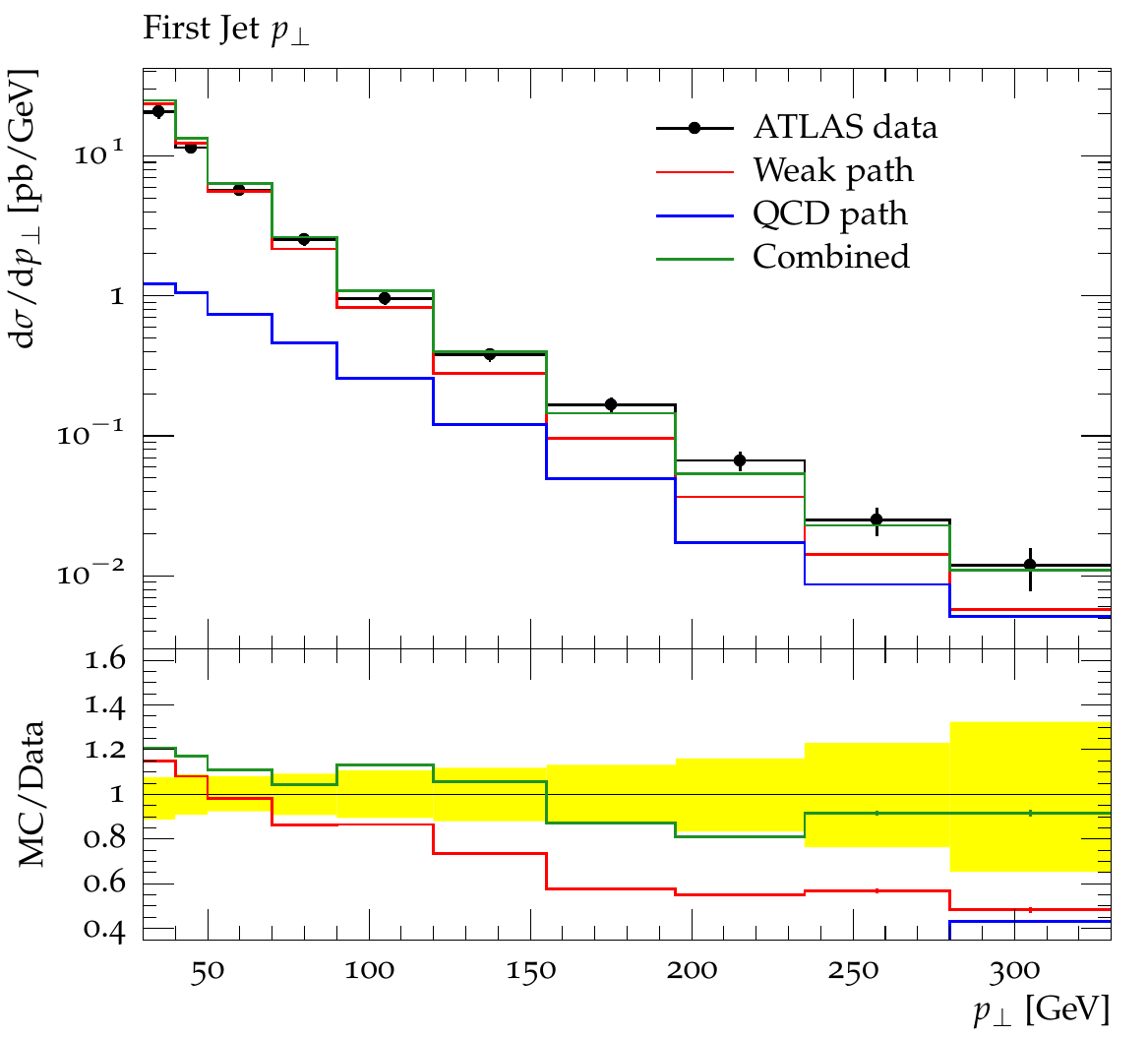}
    \includegraphics[width=0.49\textwidth]{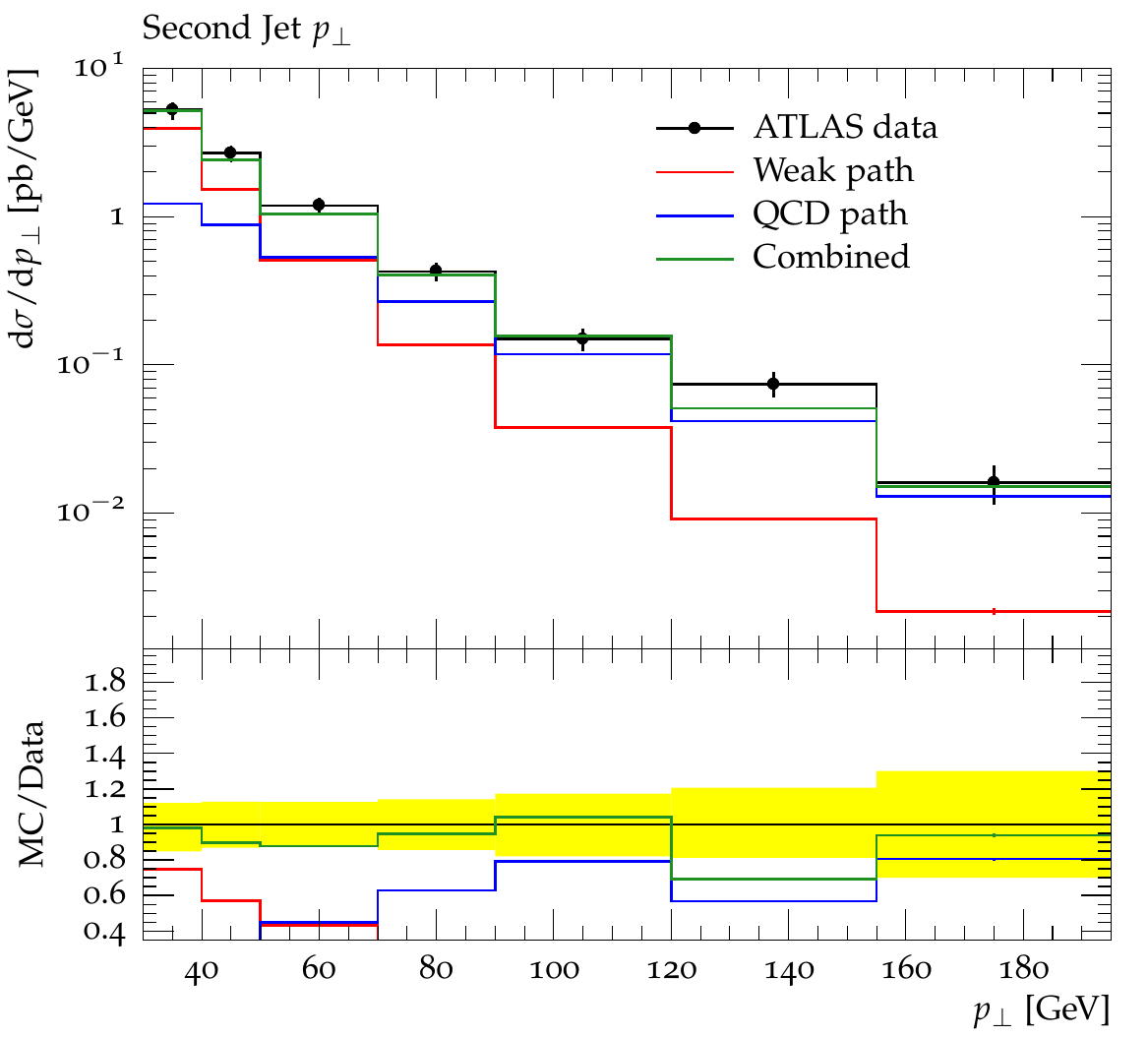}
    (a) \hspace{7.5cm} (b) \\
    \includegraphics[width=0.49\textwidth]{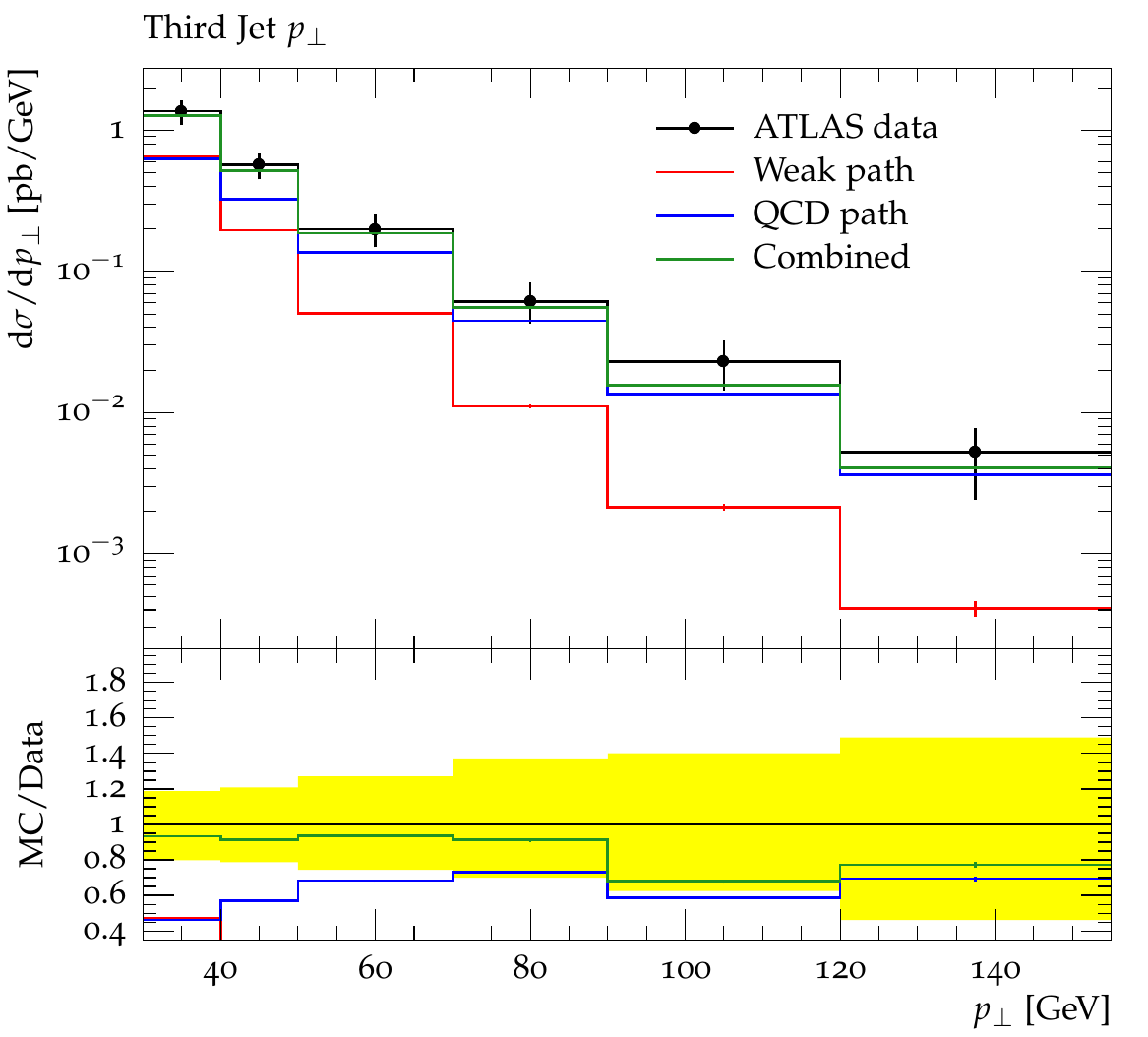} 
    \includegraphics[width=0.49\textwidth]{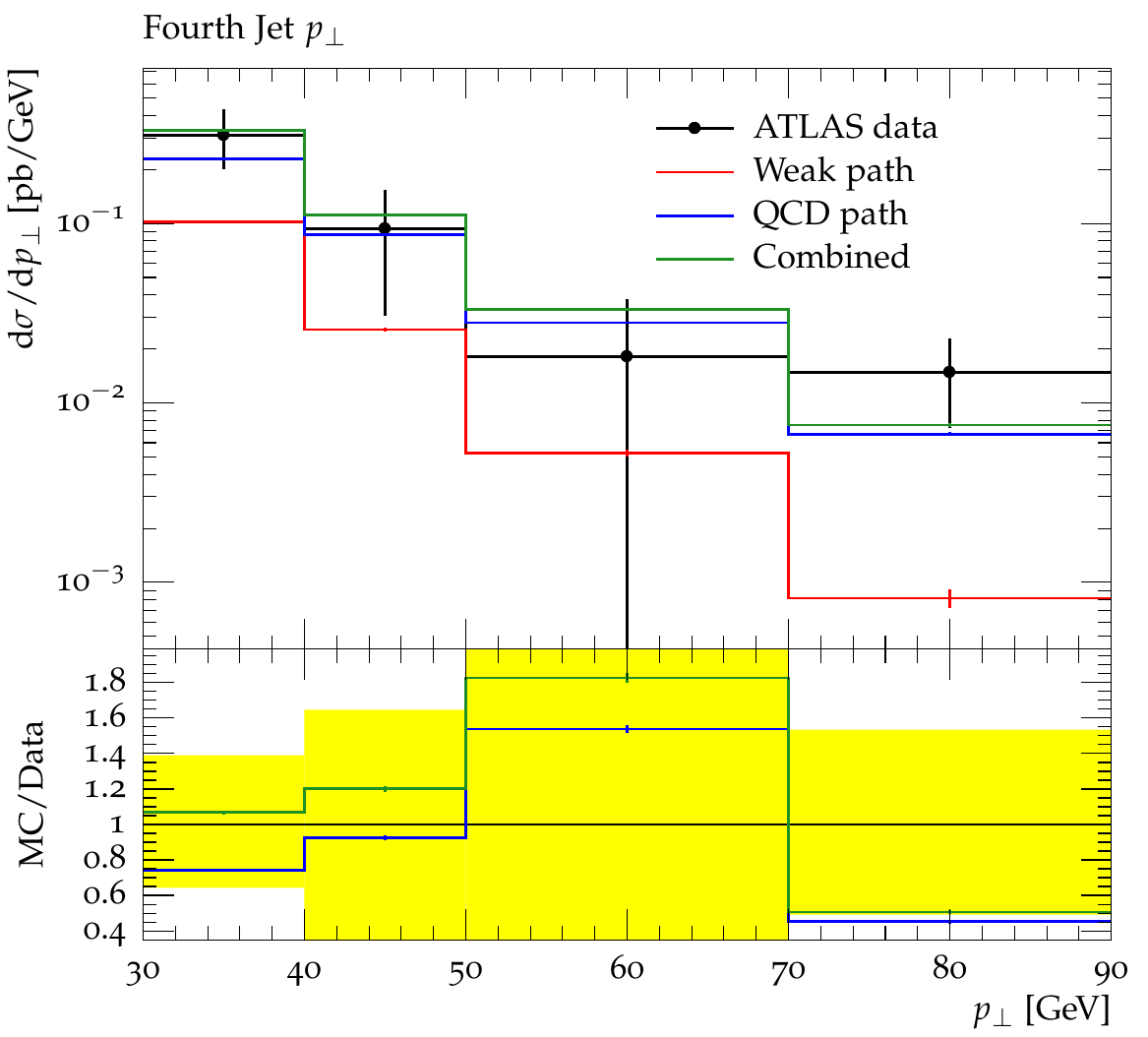} \\
    (c) \hspace{7.5cm} (d) \\
    \caption{ The $\pT$ distributions for the four hardest jets in 
    $\W$ + jet production. 
    \label{fig:wJetspT}}
    
\end{figure}

Next we turn to more exclusive quantities, beginning with the jet $\pT$ 
spectra. These are known to fall off too rapidly for the weak path alone,
Fig.~\ref{fig:wJetspT}. But here again the QCD path provides a slower
drop that nicely takes over with increasing $\pT$, giving a good overall
agreement. This is not really surprising, given that the likelihood of
emitting a $\WZ$ increases with increasing $\pT$ of the QCD process.

\begin{figure}[tbp]
  \centering
  \includegraphics[width=0.49\textwidth]{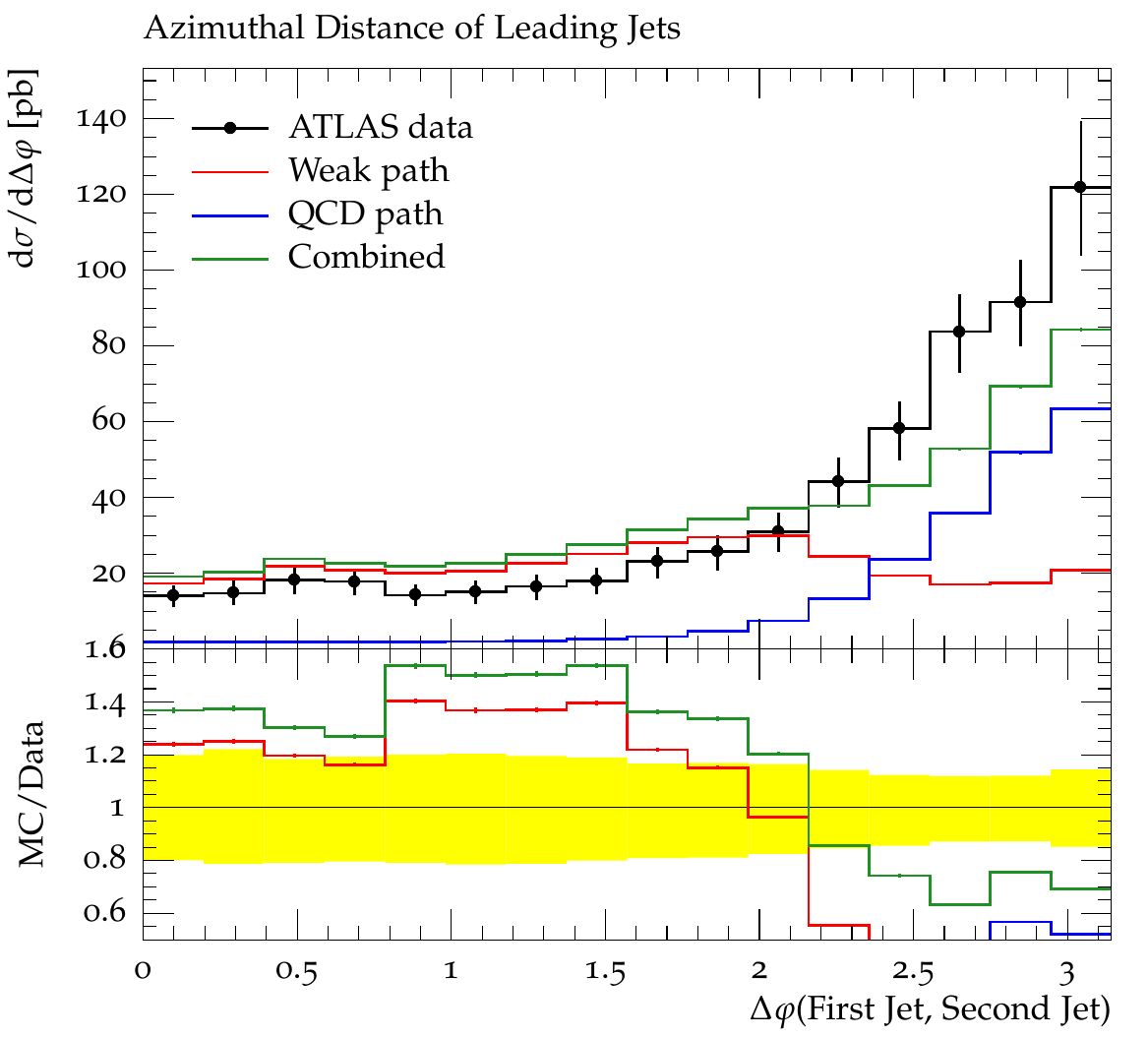} 
   \caption{ The $\Delta \varphi$ distributions for between the leading and
     sub-leading jets in $\W$ + jet production.
   \label{fig:wJetsPhi}}
\end{figure}

A further check is provided by the $\varphi$ angle between the two 
leading jets. The QCD path starts out from two
back-to-back jets, and part of that behaviour could be expected to survive the
emission of a weak boson. For the weak path the jets come from ISR, and are
therefore not expected to be particularly anticorrelated in $\varphi$.  
This is also what is observed in Fig.~\ref{fig:wJetsPhi}: the weak path is 
almost flat in 
$\Delta \varphi$, whereas the QCD path gives a clear peak around 
$\Delta \varphi = \pi$. Combining the two production channels does 
not give overwhelming agreement between data and the event generator,
however, with data having a stronger peak structure. 

\begin{figure}[tbp]
  \centering
  \includegraphics[width=0.49\textwidth]{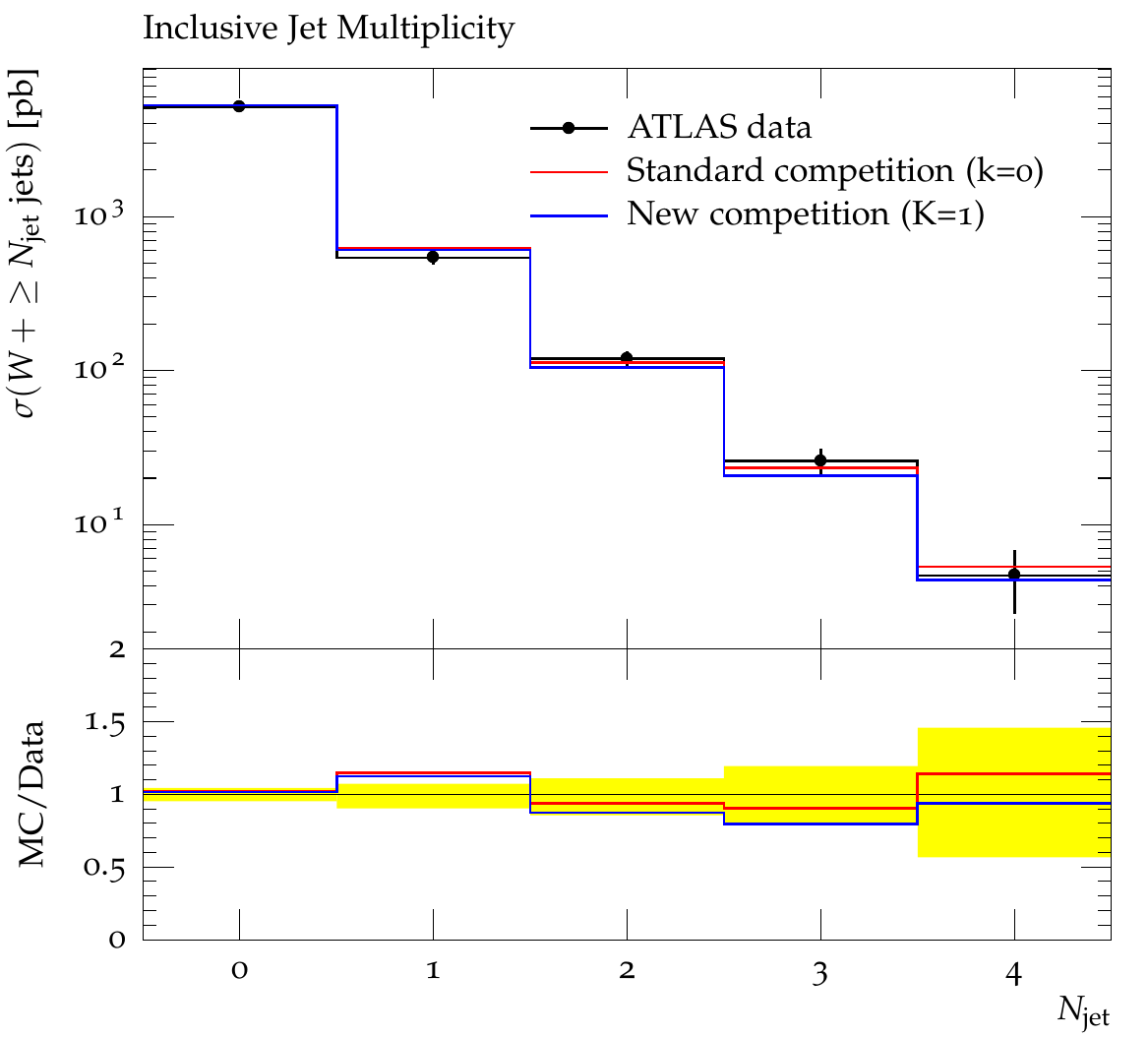}
  \includegraphics[width=0.49\textwidth]{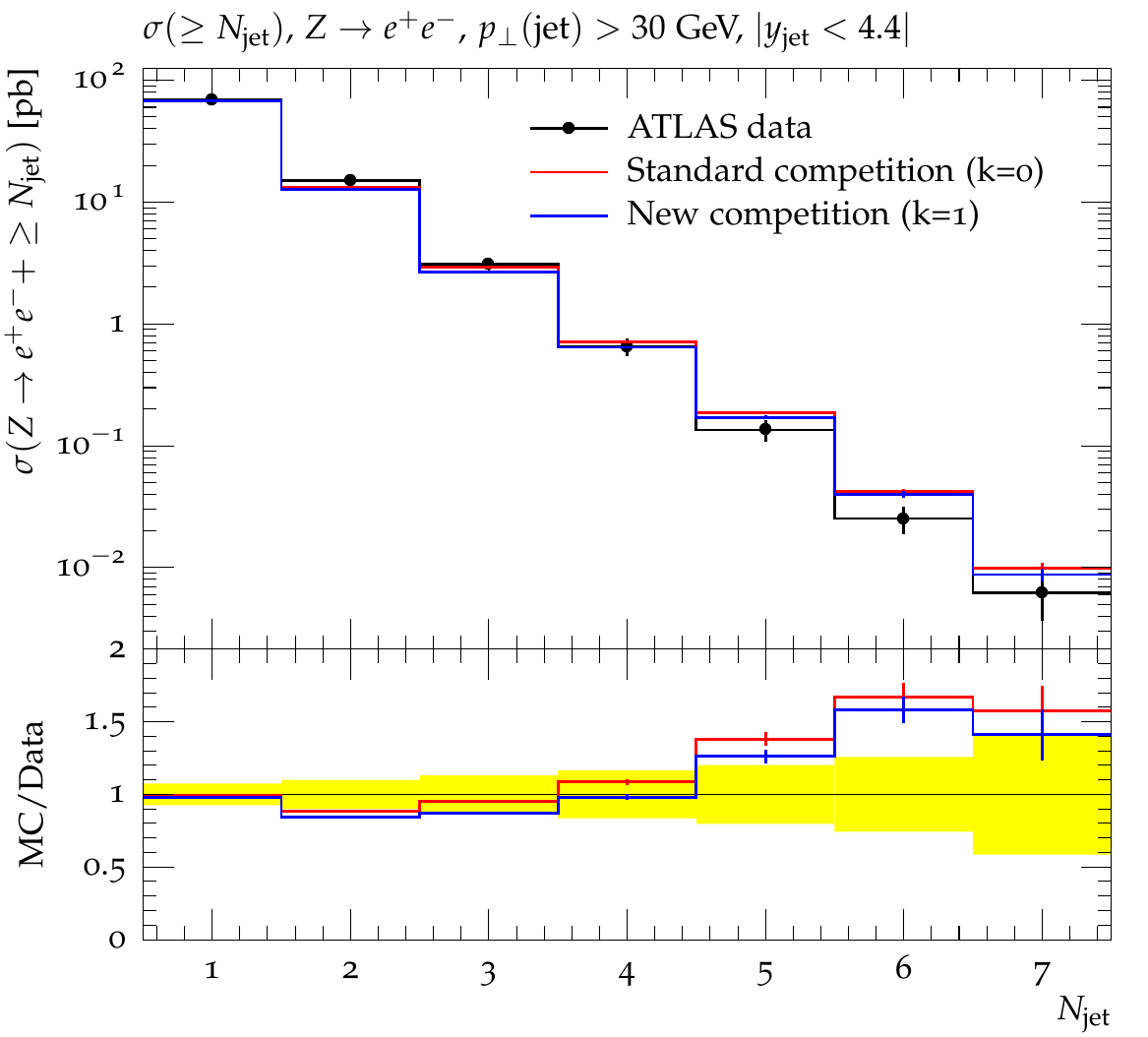}
  (a) \hspace{7.5cm} (b)
  \caption{ The inclusive jet multiplicities for two different competitions,
    (a) for $\W$ + jets and (b) for $\Z$ + jets.
    \label{fig:wJetsCompetition}
  }
\end{figure}

As mentioned previously, the implementation introduces a new parameter, $k$,
that changes the competition between the weak shower and the QCD shower. To test
whether any of the weak boson plus jets observables are sensitive to the
choice of $k$, two different simulations have been carried out. The first
simulation uses the standard $\pTe$ competition with $k = 0$ and the second
uses $k=1$, thus the weak bosons are produced earlier in the shower. 
Fig.~\ref{fig:wJetsCompetition}
shows the inclusive jet multiplicities for the two different
competitions. It may be counterintuitive that the new competition produces
a lower number of weak bosons. The explanation is that the QCD emissions can 
open new paths that allows weak emissions. Consider for instance a 
$\g \g \to \g \g$ process at a hard scale of 75~GeV. This process can
not radiate any weak bosons with the new competition, due to it
requiring the weak emissions to happen prior to any QCD
emissions (since 75~GeV is below the W mass). In the standard
competition it is possible to have a QCD emission prior to the weak emission,
thus enabling for instance a gluon splitting into two quarks followed by the
emission of a weak boson. This was also verified, by considering only those
events that had a hard scale significantly above the $\Z$ mass. And for
these, the two curves were equal within statistical uncertainties. The 
difference between the two competitions is not very large, however, and given 
the experimental uncertainty these observables do not provide any significant 
discrimination power. More differential distributions were also tested, but
none allowed a better distinction between the competitions. Thus so far it has 
not been possible to find observables that can actually tell the two 
competitions apart.

\section{Prospects for  future colliders}

The emission rate of weak bosons is expected to scale as 
$\aw \ln^2(\hat{s}/M_{\Z/\W}^2)$, and thus the effect of a weak PS is 
higher for colliders with a higher center-of-mass energy. One of the
suggestions for a possible next step beyond the LHC is a new 100~TeV
$\p\p$ collider. In this section we will redo some of the phenomenological 
studies presented in the last section, but now with the center-of-mass 
energy cranked up accordingly.

\begin{figure}[tbp]
  \centering
  \includegraphics[width=0.49\textwidth]{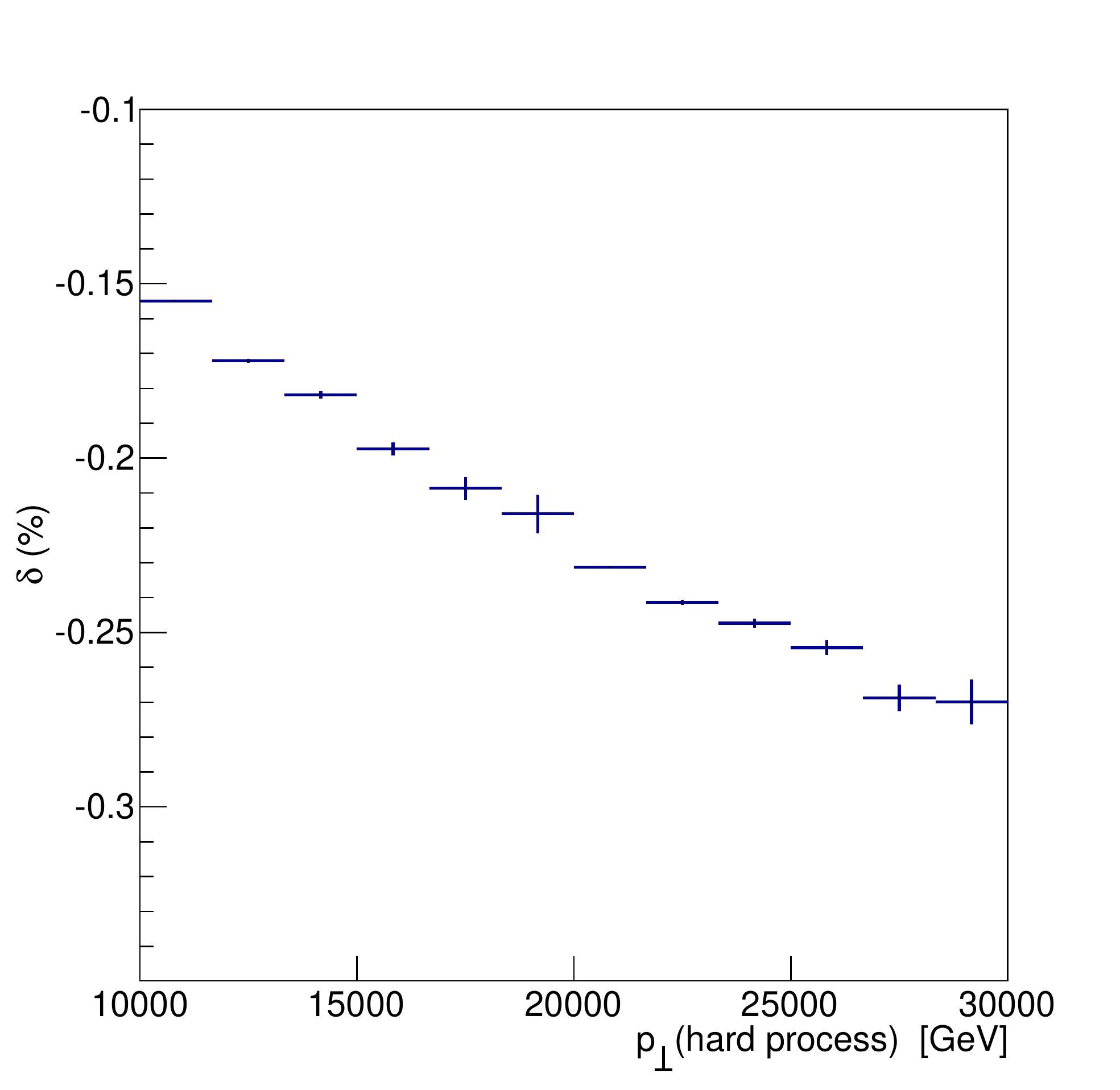}
  \caption{Weak virtual correction to di-jet production at a 100~TeV
  $\p\p$ collider, cf.\ Fig.~\ref{fig:diJetWeak}. 
  \label{fig:moretti_pT10000}
  }
\end{figure}

The weak virtual corrections to the di-jet exclusive cross section at 
100~TeV are shown in Fig.~\ref{fig:moretti_pT10000}. As before this 
equals the probability to emit at least one weak boson, up to a minus 
sign. For the di-jet with a $\pT$ around 30~TeV the corrections
reach about 30\%, a significant increase compared to the maximal 14\% 
for LHC energies. Clearly one will need to consider weak corrections 
for all processes that can have jets at large $\pT$. 

\begin{figure}[tbp]
  \centering
  \includegraphics[width=0.49\textwidth]{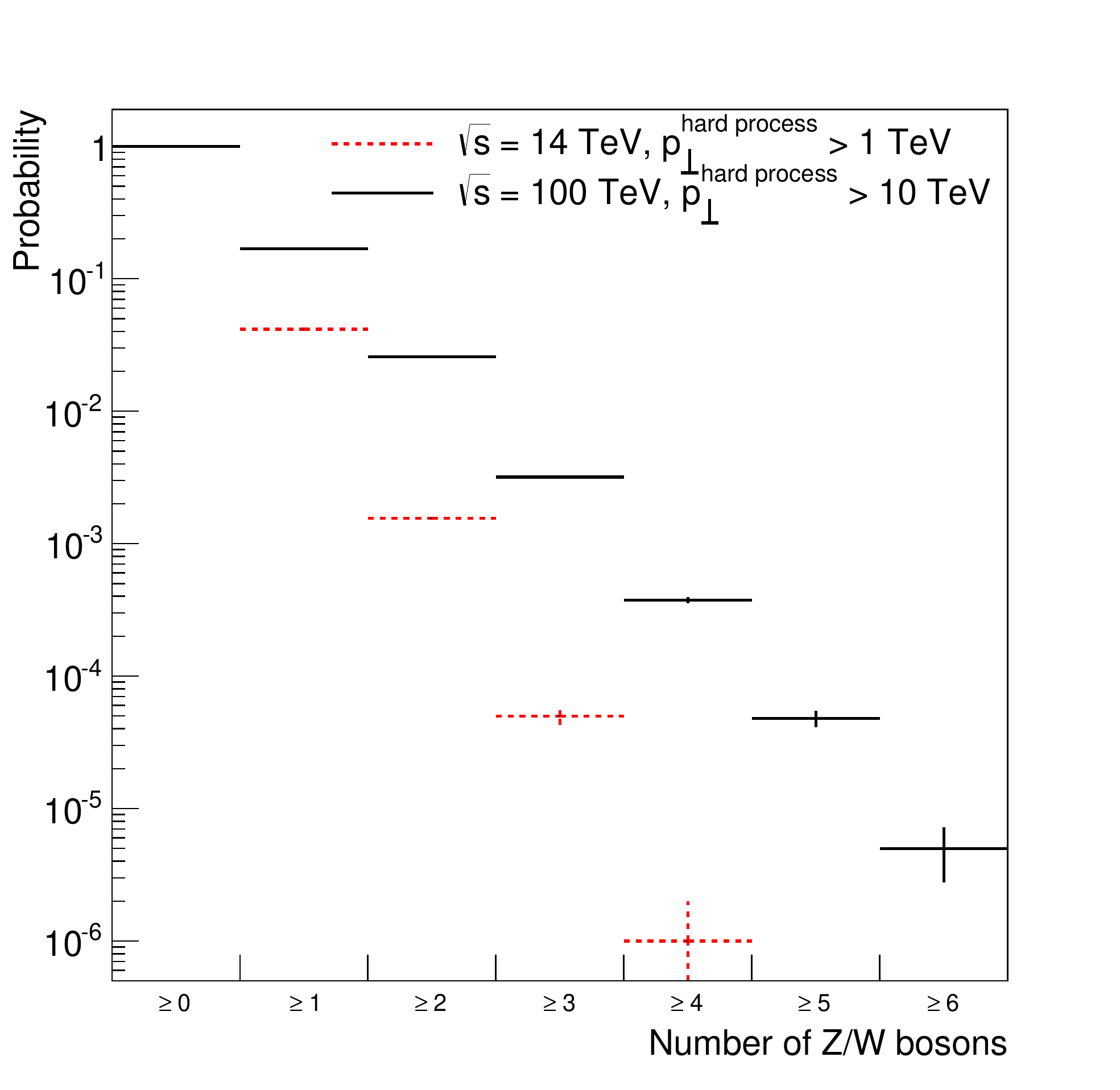}
  \includegraphics[width=0.49\textwidth]{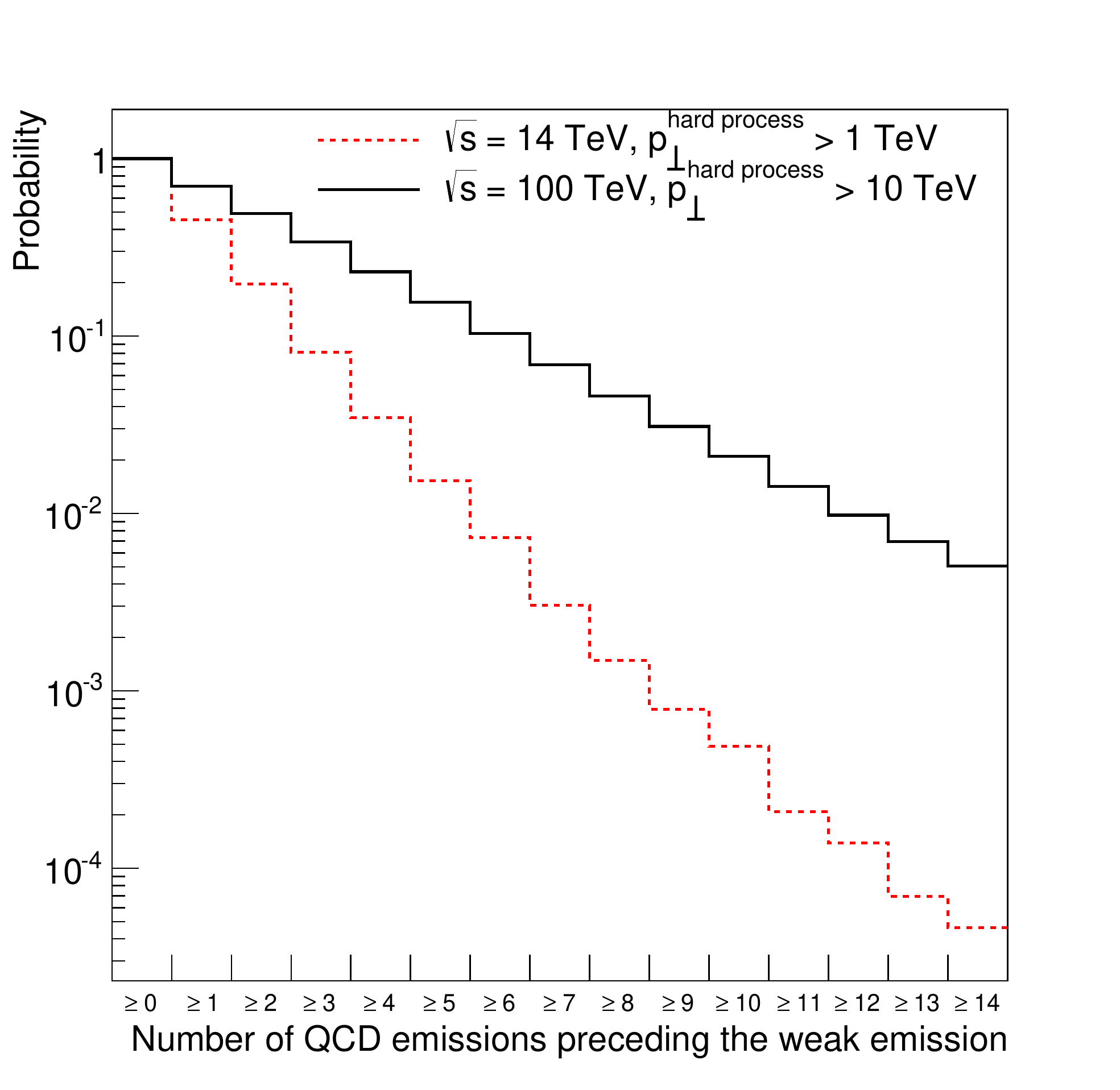}
  (a) \hspace{7.5cm} (b)
 \caption{Probability (a) for multiple emissions of weak bosons and (b) for the
    number of QCD emissions preceding the weak emission. The center of
    mass energy was set to 100~TeV and the hard process $\pT$ was above
    10~TeV. The standard competition was used.
    \label{fig:QCDvsWeakNewColl}
  }
\end{figure}

Since the emission rate for a single weak boson is enhanced 
significantly, also the rate of multiple weak emissions goes up, 
Fig.~\ref{fig:QCDvsWeakNewColl}. This is of special interest 
since currently the matching and merging schemes only describe a 
single emission of a weak boson. The probability for radiating at least 
2 weak bosons stays within a few percent for inclusive di-jet 
production. The effects may be larger for more exclusive observables. 
For instance, if you consider the production of a weak boson in 
association with jets, you would have an additional weak boson in 
$\sim$10~\% of the events (under the conditions of 
Fig.~\ref{fig:QCDvsWeakNewColl}). 

It is also interesting to note that the larger available phase space 
means that more QCD emissions can precede that of a weak boson,
Fig.~\ref{fig:QCDvsWeakNewColl}. To again obtain a one percent accuracy 
the simulations now need to include up to 11 QCD emissions before the 
weak one, which is beyond current ME capability. A matching to a shower
that can cover at least the softer $\WZ$ emissions, relative to the
large scales of the hard process, there offers obvious advantages. 

\section{Summary and outlook}

In this article we have described an implementation of weak gauge boson
emission as an integrated part of a standard parton-shower machinery,
outlined its consequences and compared it with some relevant data. 
This is a first, to the best of our knowledge.

The challenges of obtaining a realistic description have been larger than
might have been foreseen. For instance, the matching to first-order
matrix elements for $\WZ$ emission is a natural way to obtain a 
realistic description of corrections induced by the gauge boson masses, 
an issue not encountered for QCD and QED showers. A first step thus is 
to consider $\WZ$ emission off $s$-channel $2 \to 2$ QCD processes, 
where initial- and final-state radiation can be cleanly separated
(by dropping interference effects), and use these matrix elements 
to correct the shower behaviour also for other processes. 
But such a factorized description then performs rather poorly for 
$t$-channel-dominated QCD processes, necessitating a more complex 
matrix-element-correction machinery. The main drawback is that, 
in a shower language, there now arises doublecounting issues between 
what should be classified as QCD emission off a weak process and 
weak emission off a QCD process, and this has to be resolved. 
At the end of the day the weak-emission machinery therefore becomes 
more cumbersome than intended. 

Possibly the most satisfying outcome of this study is the so much
improved description of $\WZ + n$~jet data. A widespread
misconception is that showers are bound to \textit{under}estimate 
emission rates, in spite of several studies to the contrary 
\cite{Bengtsson:1986hr,Miu:1998ju,Norrbin:2000uu}. 
The poor performance of \textsc{Pythia} for $\WZ + n$~jets, with a 
clear trend to the worse for increasing $n$, has fed this myth. Now we 
see that the discrepancies essentially disappear once the possibility
of weak showers is introduced, at least within experimental errors 
and reasonable model variations, e.g.\ of $\as$. This is not to say 
that everything is perfect; as always the shower does better for 
azimuthally-averaged quantities than in more differential distributions. 

Apart from this insight, has the outcome been worth the effort? 
Not surprisingly we have shown that, even at the highest LHC 
scales, $\WZ$ emissions usually occur early in the shower evolution, such 
that the dominant $\WZ + n$~jet topologies can be generated perfectly 
well by standard matrix elements technology. So from that point 
of view the answer would be no.

However, in step with the computational advances has come the 
realization that ``raw'' order-by-order matrix elements are not enough.
Essentially all matching/merging techniques for combining the 
different fixed-$n$-jet results adopt a parton-shower perspective 
to overcome doublecounting issues. Notably a fictitious shower history 
is used to define the Sudakov form factors that are needed to turn
inclusive matrix elements into exclusive ones \cite{Catani:2001cc}. 
In the CKKW-L approach \cite{Lonnblad:2001iq} these Sudakovs are derived 
from a shower algorithm, meaning that the overall reliability of the 
matching/merging procedure is dependent on the quality of this algorithm.
Here the lack of $\WZ$ emission as a possibility can force the
adoption of less natural shower histories \cite{Lonnblad:2011xx}.
The new machinery thus opens the road to a better combined description,
even in cases when no real $\WZ$ emissions are taken from the shower 
itself. Further, the shower histories are used to reweight the fixed $\as$ 
couplings of the ME calculations to ones running as a function of 
the relevant branching scales, so also here improvements are possible.

So what lies in the future? 

Firstly, we hope that the extended shower will prove useful in its 
own right. In particular it offers a convenient tool for studying 
how the structure of jets is affected by real and virtual weak-emission 
corrections, interleaved into the standard QCD (+ QED) framework. Thus 
it is easy to get a first understanding of where effects could be 
significant, and the general order of such effects, both for jets and 
for the event as a whole. The low computer-time cost means that weak 
showers could be included routinely in event generation of any process, 
as a reminder/warning of complications that may occur, both from readily 
visible lepton pairs, from missing neutrino momenta and from the stealth 
mode of hadronically decaying weak bosons. 

Input from higher-order matrix elements will still be needed for 
precision studies. But, secondly, we have already stressed the 
improvements of matching/merging strategies that are made possible 
by including $\WZ$ emission as part of the shower evolution, so an 
obvious step is to actually upgrade the existing matching/merging 
strategies available with \textsc{Pythia}
\cite{Lonnblad:2011xx,Lonnblad:2012ng,Lonnblad:2012ix}.  

Thirdly, there are some issues that have not been addressed. One is that
we have not included the full $\gamma^*/\Z^0$ interference structure;
currently the QED machinery includes pure $\gamma^*$ effects up to some 
mass scale, while the pure $\Z^0$ kicks in above this scale. Furthermore
not all electroweak branchings are included as part of the shower,
such as $\W^{\pm} \to \W^{\pm}\gamma$, $\Z^0 \to \W^+ \W^-$ or
$\W^{\pm} \to \W^{\pm}\Z^0$. One could even imagine to include the 
Higgs in the game. However, this is on a slope of rapidly falling
shower-language relevance, so it is not clear whether the investment
in time would be worth it. 

\section*{Acknowledgements}

Work supported in part by the Swedish Research Council, contract number
621-2010-3326, and in part by the MCnetITN FP7 Marie Curie Initial 
Training Network, contract PITN-GA-2012-315877.

\end{document}